\documentclass[10pt]{article}
\usepackage{a4}
\usepackage{graphicx}
\usepackage{color}
\usepackage{epsfig}
\usepackage{amsfonts}
\usepackage{young}
\begin{document}
%\begin{large}
\title[ \textbf{AN EXACT FLUCTUATING 1/2-BPS CONFIGURATION}
\\%\end{large}\\
\begin{center}
{{\bf  Stefano Bellucci\footnote{e-mail: bellucci@lnf.infn.it }$^{,a}$
and Bhupendra Nath Tiwari\footnote{e-mail: bntiwari.iitk@gmail.com}$^{,a,b}$}}\\
\end{center}
\vspace{0.5cm}
\begin{center}
{$^a$ \it INFN- Laboratori Nazionali di Frascati,\\
 Via E. Fermi 40, 00044, Frascati, Italy}\\
\end{center}
%\vspace{0.5cm}
\begin{center}
{$^b$ \it Department of Physics, \\
Indian Institute of Technology Kanpur, \\
 Kanpur-208016, Uttar Pradesh, India}\\
\end{center}
\vspace{0.5cm}
This work explores the role of thermodynamic fluctuations in the
two parameter giant and superstar configurations characterized by
an ensemble of arbitrary liquid droplets or irregular shaped fuzzballs.
Our analysis illustrates that the chemical and state-space geometric
descriptions exhibit an intriguing set of exact pair correction functions
and the global correlation lengths. The first principle of statistical
mechanics shows that the possible canonical fluctuations may precisely
be ascertained without any approximation. Interestingly, our intrinsic
geometric study exemplifies that there exist exact fluctuating 1/2-BPS
statistical configurations which involve an ensemble of microstates
describing the liquid droplets or fuzzballs. The Gaussian fluctuations
over an equilibrium chemical and state-space configurations accomplish
a well-defined, non-degenerate, curved and regular intrinsic Riemannian
manifolds for all physically admissible domains of black hole parameters.
An explicit computation demonstrates that the underlying chemical correlations
involve ordinary summations, whilst the state-space correlations may simply
be depicted by standard polygamma functions. Our construction ascribes definite
stability character to the canonical energy fluctuations and to the counting
entropy associated with an arbitrary choice of excited boxes from an ensemble
of ample boxes constituting a variety of Young tableaux.
of string theory. \\

%\date{July 15, 2009 and, in revised form, Sept 10, 2009.}
\newpage
%%% extensive----------------------------------------------------------------------

\begin{Large} \textbf{Contents:} \end{Large}
\begin{enumerate}
\item{Introduction.}
\item{Thermodynamic Geometries.}
\item{The Fuzzballs and Liquid Droplets:}
\subitem{$3.1$\ \ Chemical Description.} \subitem{$3.2$ \ \
State-space Description.}
\item{Canonical Energy Fluctuations.}
\item{The Fluctuating Young Tableaux.}
\item{Remarks and Conclusion.}
\end{enumerate}
\section{Introduction}
Motivated from the microscopic understanding of black holes in string
theory, brane configuration and microscopic proposals \cite{hep-th/0508023,
hep-th/0107119}, we focus on investigating the possibilities of a covariant
thermodynamic geometric study of a class of coarse grained configurations.
These solutions are described in terms of the chemical parameters and a set
of arbitrarily  excited boxes of the random Young tableaux. In particular,
we intend to study the thermodynamic geometry arising from the canonical
energy and counting degeneracy for the two parameter finite temperature
solutions. Ref. \cite{07073601v2} offers an interesting motivation for
the case of near-extremal black holes in $AdS_5$. The main idea is first
to develop a geometric notion for black hole thermodynamics and then to
relate it with the existing microscopic quantities already known from CFT
configurations \cite{hep-th/0508023, hep-th/0107119}. Furthermore,
we incorporate all order fluctuations in the exact canonical energy
\cite{SM} and entropy associated with the counting of excited boxes
in random Young tableaux and analyze their respective contributions
to the covariant thermodynamic geometries.
The phenomenon of the statistical fluctuation in the bubbling solutions
involves the same asymptotic charges as that of the considered giant black holes.
To be precise, we consider an ensemble of horizonless geometries and the black hole
arises as a coarse graining of such states.

Lunin and Mathur have connoted a conjecture \cite{lu-ma} that the
black hole microstates may be characterized by string theory
backgrounds with no horizons. Moreover, these solutions saturate
an exact bound on their angular momenta, and have the same set of
asymptotic charges as the respective black hole whose horizon area
classically vanishes. Some finite temperature supersymmetric \cite{0411001v2}
and associated non-supersymmetric generalizations appear as well,
which correspond to definite black holes with finite size horizons. 
As a consequence, we are interested in finding state-spaces for the
BPS black holes which are already known to have a finite size horizon
at zero temperature \cite{mathur1, mathur2}. Furthermore, the complete
class of two parameter\footnote{The two parameters of the interest
are the charge and effective canonical temperature of the solution
\cite{hep-th/0508023, hep-th/0107119}. It is worth mentioning that 
this temperature is not the physical temperature of the black holes.
However, it arises as the canonical averaging e.g. Eqn(\ref{effTemp}).}
supersymmetric black hole solutions in five dimensions are well-known
since the invention of \cite{LLM}. Ref. \cite{09103225v2} brings out
an interesting issue of the horizonless solutions in AdS space and thereby
constructs its relation for the case of the small black hole solutions.
The associated finite temperature solutions \cite{0411001v2,09103225v2}
could be thought as a deviation from the extremality condition. Importantly,
the near-extremal configurations \cite{07073601v2} may further indicate a
set of interesting issues for the present discussion thermodynamic geometry. 
Thus, it would be inspiring to analyze the thermodynamic intrinsic manifold
for the giants and superstars configurations having two parameters, which
characterize fluctuation about an ensemble of equilibrium CFT microstates
\cite{hep-th/0508023, hep-th/0107119}. From the perspective of the thermodynamic
geometry, this sets a good motivation to study the issue of statistical
correlations for the black holes in string theory.

Interestingly, the existence of such a geometric structure in
equilibrium thermodynamics was introduced by Weinhold \cite{wien1,wien2}
through an inner product in the space of equilibrium chemical
potentials defined by the minima of the internal energy function
$U=U( \lambda_i, V, S, T)$ as the Hessian function
$h_{ij}= \partial_i \partial_j U $.  In this description, the
quantities $ \lbrace \lambda_i, V, S, T \rbrace $ are defined as
the chemical potentials $ \lambda_i$, phase-space volume $V$,
entropy $S$ and the temperature $T$ of the underlying equilibrium
statistical configuration. In order to provide a physical scale,
we need to restrict negative eigenvalues of the metric tensor,
which may be accomplished by requiring that the volume or any other
external parameter of the given fluctuating giant configurations
be held fixed.

Let us recall that the Riemannian geometric structure thus
involved indicates a certain physical significance ascribed to
giants and dual giants. The associated inner product structure
on the intrinsic space may either be formulated in the entropy
representation, as a negative Hessian matrix of the counting
entropy with respect to the extensive variables, where the total
number of boxes and number of excited boxes in a random Young
Tableaux define the degeneracy, or be formulated in the energy
representation, as the Hessian matrix of the associated energy
with respect to the intensive chemical potentials
\cite{hep-th/0508023,hep-th/0107119}. Such geometric considerations
have been studied earlier in the literature, as the covariant
thermodynamics whose application in the energy representation
has first been considered by Weinhold \cite{wien1,wien2}. Whilst,
the entropy representation in which we shall investigate fluctuating
Young tableaux has been considered in the view-points of thermodynamic
fluctuation theory \cite{RuppeinerRMP,math-ph/0507026,RuppeinerA20,
RuppeinerPRL,RuppeinerA27,RuppeinerA41}.

The present paper is thus intended to analyze both pictures,
and indeed our study strikingly provides a set of exact covariant
thermodynamic geometric quantities ascribed to the giants and to the
superstars \cite{hep-th/0508023}. Recall that the thermodynamic
configuration of the $1/2$-BPS (dual) giants may be parameterized
by the underlying chemical potentials, $\lambda_1, \lambda_2$.
Thus, fluctuations in the associated ensemble may be described by
the minima of the energy function $E= E(\lambda_1, \lambda_2)$.
Explicitly, the Weinhold metric tensor of an intrinsic Riemannian
space spanned by the chemical potentials may be given as $g_{ij}=
\partial_i \partial_j E(\lambda_1, \lambda_2)$ and turns out to be
conformal to the Ruppenier metric\footnote{The present paper
assumes that the terms (Ruppenier, Weinhold) geometry and
(State-space, Chemical) geometry are synonyms and can be used
interchangeably. It is worth mentioning that the geometry associated
with the Weinhold metric tensor is said to be the chemical geometry
and the other whose metric tensor is defined by the Ruppenier metric
is said to be the state-space geometry. The general definitions are
given by the Eqns(\ref{Wienmetricgen},\ref{Ruppmetricgen}). Subsequently,
The Eqns(\ref{Wienmetric},\ref{Ruppmetric}) explain the precise
consideration of the analysis of the present paper.}
with the temperature as the conformal factor. Thus, the metric as an inner
product on the intrinsic surface of chemical potentials is necessarily
ensured to be symmetric, positive definite and satisfies the triangle inequality,
since the energy function has a minimum configuration in the equilibrium.

As a matter of fact, a few simple manipulations illustrate that
the conformally related state-space curvature is inversely
proportional to the singular part of the free energy associated
with long range correlation(s), whose regularity consequently
signifies the fact that underlying giant configurations have no
phase transition(s) over the range of admissible boxes. It is
worth to mention that the Ruppeiner formalism, which has further
been applied to diverse condensed matter systems with two dimensional
intrinsic Riemannian spaces have been found to be completely
consistent with the scaling and hyper scaling relations involving
certain critical phenomena and in turn reproduce their corresponding
critical indices, see for review \cite{RuppeinerRMP,math-ph/0507026}.
In general, such geometric notions remain well-defined and correspond
to an interacting statistical basis for diverse extremal and non-extremal
black brane configurations \cite{bnt,SST,0606084v1}.

There exists a large class of black holes which have the
spherical horizon topology in the physical regime, whose thermodynamic
configurations are quite energizing in their own respect, as we shall exhibit
in the next section. Subsequently, we shall analyze the behavior
of collective state-space correlations, and in particular, we wish to
illustrate certain peculiarities of the state-space geometry which in
the large charge limit provides the correlation length for the associated
microscopic configurations. 
The purpose of this paper is to investigate the relation between
the covariant thermodynamic geometries and the microscopic counting
perspective from the total and excited boxes of random Young tableaux.
Both of these notions have received considerable recent attention from
the perspective of black hole physics and the corresponding thermodynamics.
Thus, the goal here is to explore underlying microscopic arguments behind
the macroscopic notion of the covariant thermodynamic geometries arising
from the energy/ degeneracy of fluctuating (dual) giant configurations.
Furthermore, the underlying state-space curvature for higher dimensional
black holes appears to be intertwined with vacuum fluctuations existing
in the possible brane configurations, and thus one may revel certain
intriguing informations of the possible microscopic CFTs \cite{fuzzball}.

In this connection,
what follows here shows that the thermodynamic intrinsic geometries and
their respective microscopic resolutions may however be pursued by
an intriguing involvement of the framework of Mathur's fuzzball
solutions \cite{LLM,fuzzball,LM}. Such a perspective thus discloses
an indispensable ground with the statements that the thermodynamic
scalar curvatures of interest not only has exiguous microscopic
knowledge of the black hole configuration, but also it has rich
intrinsic Riemannian geometric structures. Specifically, we procure
that the configurations under the present analysis are effectively
attractive in general, while they are stable only if at least one
of the parameters remain fixed. Our hope is that finding statistical
mechanical models with like behavior might yield further insight
into the microscopic properties of black holes and a conclusive
physical interpretation of their state-space curvatures and
related intrinsic geometric invariants.

Within the well understood framework of the microscopic counting technique,
what fundamentally follows is that the present geometric notions
implicitly assume an infinite set of subensembles of the most
general statistical system having an integrated ensemble \cite{LM}.
We thus explore the intriguing role of chemical fluctuations in
the united canonical energy, or box counting statistical entropy,
of the underlying black brane configurations. Consequently, we
find that the associated intrinsic Riemannian geometries being
Legendre transforms of each other, indeed describe corresponding
fluctuations of the excited droplets or fuzzballs. For example,
the canonical energy corrected by the contributions coming from
the fluctuations of an ensemble of excited boxes may be regarded
as a closer approximation to that of the simple energy in the
corresponding subensembles. We shall show that such considerations
apply as well to two parameter fuzzballs \cite{LLM,fuzzball,LM}
or the liquid droplets \cite{liquid}, when considered as an union
subensembles. The specific forms of the infinite summation and
special polynomial contributions may thus be calculated for a
class of supergravity black holes \cite{LLM}.

It is noteworthy that the applicability of this analysis
presupposes that the underlying net ensemble is thermodynamically
stable, which requires a positive specific heat, or correspondingly
that the Hessian matrix of the energy function or the negative Hessian
matrix of the counting entropy, be positive definite. The energy
function and degeneracy for the (dual) giant or superstar systems
incorporating such contributions \cite{LLM,fuzzball,LM,liquid} may
easily be written to an appropriate form, without any approximation.
The present article discovers that such contributions to the canonical
energy/ box counting entropy induce certain bumps in the scalar curvature
of underlying intrinsic thermodynamic surfaces spanned by the chemical
potentials, or the total and excited boxes of the Young tableaux.

The present paper is organized as follows. In the first section,
we have presented the motivations to study the intrinsic Riemannian
surfaces obtained from the Gaussian fluctuations of the giants or
superstars canonical energy or counting entropy. In particular,
we have outlined some microscopic implications of the thermodynamic
geometry for the giants and superstars, in the lieu of the underlying
microscopic configurations, and the remaining significations have
been summarized in the final section. In section $2$, we shall
briefly explain what are the intrinsic thermodynamic geometries
based on a large number of equilibrium parameters characterizing
the giant and superstar black hole configurations. In section $3$,
we obtain the two parameter giants (superstars) canonical
energy and counting entropy as a function of chemical potentials
and in terms of the number of excited and total boxes of the Young
tableaux. In section $4$, we investigate the chemical geometry
for the giant (/superstar) black holes, and examine two parameter
fluctuating canonical configurations.

In section $5$, we focus our attention on the state-space geometry
of the aforementioned configurations, with some excited boxes and
an arbitrarily random Young tableaux. Explicitly, we investigate
the nature of the state-space geometry thus defined, as an intrinsic
Riemannian manifold obtained from the entropies, with all possible
contributions being considered in the most general canonical ensemble.
We have explained that such an intrinsic geometric configuration,
obtained from an effective canonical energy or degeneracy of the
giants/ superstars, results to be well-defined and pertains to an
interacting statistical system. Finally, section $6$ contains a set
of concluding issues and a possible discussion of the thermodynamic
chemical and state-space geometries. The general implications thus
obtained may thus divulge the geometry of both the chemical and the
equilibrium microstate acquisitions which are the matter of an AdS/CFT
correspondence. Furthermore, our results are in close connection
with the microscopic implications arising from the underlying CFTs,
and thus offer definite physically sound explanations of the finite
chemical potential/ finitely many excited boxes configurations.
\section{Thermodynamic Geometries}
The present section presents a brief review of the essential
features of thermodynamic geometries from the perspective of the
application to the two parameter configuration.%: giants and superstars. 
As mentioned in the introduction, in order to illustrate the
connection between thermodynamic aspects and microscopic counting,
we focus our attention on the geometric nature of underlying
fluctuating chemical parameters and number of excited (unexcited)
boxes in an ensemble of finitely  random Young tableaux. The respective
analysis is divulged in the neighborhood of small chemical potentials
or in an arbitrarily large number of excited boxes. In the present
context, these limits essentially provide an equilibrium configuration.
In the next section, we shall motivate the thermodynamic geometric
nature  for the (dual) giants and superstars.

In order to illustrate the basic notion of thermodynamic fluctuations,
let us consider an intrinsic Riemannian geometric model, whose
covariant metric tensor may be defined as the Hessian matrix of the
canonical energy, with respect to a finite number of arbitrary
chemical potentials carried by the giant gravitons and superstar,
considered in a given equilibrium configuration at fixed volume, or
any other parameters of the solution. Specifically, let us define a
representation $ E( \lambda_i, T) $ for given canonical energy, chemical
potentials, and temperature $ \lbrace E,  \lambda_i, T \rbrace $.
In the present consideration, the energy, as defined below, may
further be shown to be solely a function of the chemical potentials,
which in fact accompanies an intriguing set of expressions, as we have
indicated in general by Eqs. \ref{energy}, \ref{entropy}. The underlying
method, in general, results to be in an intrinsic Riemannian geometric
configuration, which is solely based on nothing but the giant
chemical potentials. Such a general consideration in fact yields an
intrinsic space spanned by the $n$ chemical potentials of
the theory under examination, and thus it exhibits a $n$-dimensional
intrinsic Riemannian manifold $ M_n $. The components of the
covariant metric tensor of the so called chemical geometry
\cite{wien1,wien2} may be defined as
\begin{scriptsize}
\begin{eqnarray} \label{Wienmetricgen}
g_{ij}:=\frac{\partial^2 E(\vec{x})}{\partial x^j \partial x^i}
\end{eqnarray}
\end{scriptsize}
where the vector $\vec{x} =(\lambda_i, V, S, T) \in M_n $.
As anticipated in the appendix A, it is worth to mention that
the thermodynamic curvature corresponds to the nature of the intrinsic
correlation present in the statistical system. This will altogether
imply that the scalar curvature for two component systems can be
thought of as the square of the correlation length, at some given
non-zero configuration temperature, to be $R( \lambda_1,\lambda_2)
\sim \chi^2 $, where we may identify the $\chi(\lambda_1,\lambda_2)$
to be the correlation length of the concerned chemical system.
Consecutively, it is not surprising that the geometric analysis,
based on the Gaussian approximation to the classical fluctuation
theory, precisely yields the correlation volume of the chosen configuration.
This strongly suggests that, even in the context of chemical reactions
or in any closed system, a non-zero scalar curvature might provide
useful information regarding the range of microscopic phase-space
correlations between various components of the underlying statistical
configuration.

Note that such Riemannian structures, defined by the metric tensor
in a chosen representation, are in fact closely related to the
classical thermodynamic fluctuation theory \cite{RuppeinerA20,
RuppeinerPRL, RuppeinerA27, RuppeinerA41} and existing critical
phenomena. The probability distribution $\Omega(x)$ of thermodynamic
fluctuations over an equilibrium intrinsic surface naturally
characterizes an invariant interval of the corresponding
thermodynamic geometry, which in the Gaussian approximation reads
\begin{scriptsize}
\begin{eqnarray}
\Omega(x)= A \ \ exp [-\frac {1}{2} g_{ij}(x) dx^i \otimes dx^j],
\end{eqnarray}
\end{scriptsize}
where the pre-factor $A$ is the normalization constant of the
Gaussian distribution and $\otimes$ denote a symmetric product.
The associated inverse metric may easily be shown to be the second
moment of the fluctuations or the pair correlation functions, and
thus it may be given as $g^{ij}= < x^i \vert x^j>$, where 
$\lbrace x_i \rbrace $'s are the intensive chemical variables
conjugated to the charges $\lbrace X^i \rbrace$ of the Legendre
transformed entropy representation. Moreover, such Riemannian
structures may further be expressed in terms of a suitable
thermodynamic potential, obtained by certain Legendre transforms,
which correspond to certain general coordinate transformations
on the equilibrium thermodynamic manifold.

Following \cite{RuppeinerRMP, 0606084v1, 0510139v3}, it turns out
that the natural inner product on the state-space manifold may
easily be ascertained for an arbitrary finite parameter black
brane configuration, and the concerned state-space turns out to
be a $n$-dimensional intrinsic manifold $ M_n $. Typically, the
associated entropy $S(X^i)$ as an embedding function defines
the covariant components of the metric tensor of the thermodynamic
state-space geometry, which has originally been anticipated by
Ruppeiner in the related articles \cite{RuppeinerRMP,RuppeinerA20,
RuppeinerPRL, RuppeinerA27, RuppeinerA41}. Here, we shall take
this representation of the intrinsic geometry, and thus find
that the covariant components of state-space metric tensor
may be defined to be
\begin{scriptsize} 
\begin{eqnarray} \label{Ruppmetricgen}
g_{ij}:=-\frac{\partial^2 S(\vec{X})}{\partial X^i \partial X^j}
\end{eqnarray}
\end{scriptsize}
We may thus explicitly describe the present state-space geometric
quantities as simply an intrinsic two dimensional Riemannian
manifold for $1/2$-BPS configurations. Furthermore, the underlying
state-space geometry may parametrically be defined by the two invariant
parameters, {\it viz.}, $\vec{X}= (n,M) \in M_2$.
We may therefore notice that the components of the state-space
metric tensor are related to the statistical pair correlation
functions, which may as well be defined in terms of the parameters
describing the dual microscopic conformal field theory living on the
boundary. This is because of the fact that the underlying metric
tensor comprising Gaussian fluctuations of the entropy defines the
state-space manifold for the rotating black brane configuration.
We may thus easily perceive, in the present consideration, that the
local stability of the underlying statistical configuration
requires that the principle components of state-space metric
tensor $\{ g_{ii} \ | \ i:= n, M \}$ signifying heat capacities
should be positive definite
\begin{scriptsize}
\begin{eqnarray}
g_{nn}(n,M) &>& 0 \nonumber \\
g_{MM}(n,M) &>& 0
\end{eqnarray}
\end{scriptsize}
Moreover, the positivity of the state-space metric tensor imposes
a stability condition on the Gaussian fluctuations of the underlying
statistical configuration, which requires that the determinant and
hyper-determinant of the metric tensor must be positive definite.
In order to have a positive definite metric tensor $\Vert g(n,M)\Vert$
on the two dimensional state-space geometry, one thus demands that
the determinant of metric tensor must satisfy $\Vert g \Vert >0$,
which in turn defines a positive definite volume form on the
concerned state-space manifold. Furthermore, it is not difficult
to calculate the Christoffel connection $\Gamma_{ijk}$, Riemann
curvature tensor $R_{ijkl}$, Ricci tensor $R_{ij}$, and the scalar
curvature $ R $ for the two dimensional state-space intrinsic
Reimannian manifold $(M_2,g)$. Remarkably, it turns out that
the above two dimensional state-space scalar curvature appears
as the inverse exponent of the inner product defining the pair
correlation functions between arbitrary two equilibrium microstates
characterizing the black brane statistical configuration. 

Notice that there exists an intriguing relation of the scalar
curvature of the state-space intrinsic Riemannian geometry,
characterized by the parameters of the equilibrium microstates,
with the correlation volume of the corresponding black brane
phase-space configuration. Additionally, Ruppeiner has revived 
the subject with the fact that the state-space scalar curvature
remains proportional to the correlation volume $\tilde{\chi}^d$,
where $d$ is the system spatial dimensionality and $\tilde{\chi}(n,M)$
is its correlation length, which reveals related information residing
in the microscopic models \cite{RuppeinerA20}. It is worth mentioning
further that the state-space scalar curvature in general signifies
possible interaction in the underlying statistical configuration. 
Furthermore, one may appreciate that the general coordinate
transformations on the state-space manifold thus considered
expound to certain microscopic duality relations associated with
the fundamental invariant charges of the configurations.

From the perspective of an intrinsic Riemannian geometry, it seems
that there exists an obvious mechanism on the black brane side, and
that it would be interesting to illuminate an associated stringy notion
for the statistical correlations to the microstates of the giant and
superstar solutions or vice-versa. In this concern, it turns out
that the state-space constructions so described might elucidate
certain fundamental issues, such as statistical interactions and
stability of the underlying brane configurations with spins and
non-equal $R$ charges. Nevertheless, one may arrive to a definite
possible realization of the equilibrium statistical structures,
which is possible to determine in terms of the parameters of an
ensemble of microstates describing the equilibrium configurations.

The relation of a non-zero scalar curvature with an underlying
interacting statistical system remains valid even for higher
dimensional intrinsic Riemannian manifolds, and the connection of a
divergent scalar curvature with phase transitions may accordingly
be divulged from the Hessian matrix of the considered energy/
counting entropy. It is significant to remark that our analysis
takes an intriguing account of the scales that are larger than the
correlation length and considers that only a few microstates do not
dominate the whole macroscopic equilibrium intrinsic quantities.
Specifically, we shall focus on the interpretation that the
underlying energy includes all order contributions from a large
number of subensembles of the fluctuating microstates, and thus
it characterizes our description of the geometric thermodynamics
for (dual) giants, superstars and fuzzballs.

Our geometric formulations thus tacitly involves an unified
statistical basis, in terms of the chosen union of subensembles.
Although the analysis has only been considered in the limit of
small fluctuations, however the underlying correlation length takes
an intriguing account upon the quartic corrections of the canonical
energy or the counting entropy. With this general introduction to
the thermodynamic geometries defined as the (negative) Hessian
function of the canonical energy (or counting entropy), let us now
proceed to investigate the energy/ entropy of two parameter giants,
superstars and their thermo-geometric structures. In the present
investigation, we shall focus our attention on the interpretation
that the underlying energy includes contributions from a large number
of excited boxes, and thus our description of the geometric thermodynamics
extends itself to the spinless $1/2$ BPS supergravity configurations.
\section{The Fuzzballs and Liquid Droplets}
In this section, we shall first review the giant and superstar
configurations \cite{hep-th/0508023, hep-th/0107119} and subsequently
compute the canonical energy and box counting degeneracy in the
required form. Thereafter, we explore the thermodynamic geometries of
the giants and superstars arising from the consideration of type IIB
string theory. For obtaining the energy as a function of an effective
temperature and chemical potential, we consider the liquid droplet model
of giants and superstars. The present paper considers, {\it viz.} (i)
the canonical energy with two distinct parameters $ (T, \lambda)$,
with $T$ as an effective canonical temperature and $\lambda$ as
the chemical potential dual to the underlying $R$-charge of the theory
and (ii) the box counting entropy with two distinct large integers
$(n,M)$, ($n$ corresponds to the number of excited boxes and $M^2$
corresponds to the total number of possible boxes). In order to obtain
the desired expression for the energy and counting entropy, we consider
the dual $\mathcal{N}=4$ super Yang Mills theory on $AdS_5 \times S^5$
for type IIB string theory.
\subsection{Chemical Description}
In order to understand the very basic picture of the giant and
superstar configurations \cite{hep-th/0508023, hep-th/0107119},
we shall restrict ourselves to the field theory description defined
on $ \mathrm{R} \times S^3$ and focus our attention by recalling the
$1/2$ BPS sector of $\mathcal{N}=4 \ SYM$ with conformal dimension
$\Delta=R$. It is known \cite{hep-th/0508023, hep-th/0107119} that the
isomorphism group involves an $SO(6)$ symmetry, and thus we shall consider
an embedding $SO(6) \subset SO(2) \times SO(2) \times  SO(2)$, which may in 
general have various states $\{ X, \overline{X}, Y, \overline{Y},Z, \overline{Z} \}$
corresponding to either of the $SO(2)$. However, for spinless black holes,
as there is no spin structure in $SO(2)$, one obtains $ J=0=T$. Thus an
appropriate dimensional reduction of the above configuration yields to a
simple quantum mechanical system involving only $\{ X, \overline{X} \}$+
a bunch of BPS states. Following \cite{0111222, 0403110}, the problem under
consideration may thus be mapped to a real matrix model with a real orthogonal
matrix $X$.

The problem under consideration may be viewed as a $N$ fermion configuration
lying in a one dimensional harmonic oscillator. One may in principle arrive
at the solution and thus find eigenvalues and eigen vectors of the
equations of motion. For the present purpose, let the eigen values of
the real matrix wave function be $\{\lambda_i\}_{i=1}^N$, then the
complete problem may be described by considering the Van der monde determinant
of the function $X$. Surprisingly, one recognizes that the product
$\prod_{k+j} (Tr X^k)^j$ may be expressed in terms of the eigen
values $\{\lambda_i\}$. The fermions under consideration thus
acquire a ``Fermi sea'', which may physically be understood as
follows. We observe that the individual energies $\epsilon_i= e_i
\hbar + \hbar /2$ acquire a gap of
\begin{scriptsize}
\begin{eqnarray}
r_i&=& \frac{1}{\hbar} (E_i-E_i^g)= e_i-i+1, \ \forall \
i=1,2,..., N
\end{eqnarray}
\end{scriptsize}
In order to explicate the dual gravity picture, we shall consider the
well-known LLM description, see for details \cite{LLM}. It thus
follows from type $IIB$ string theory that the classical moduli
may be described with a symmetric $AdS_5 \times S^5$ background,
which has the symmetry of $\mathrm{R} \times SO(4) \times SO(4)
\times U(1)$, having generators $ \partial_t $, $ S^3 \subset AdS_5
$, $\widetilde{S}^3 \subset S^3$ and $ \partial_{\phi}$, respectively,
for the concerned symmetry components. It has been pointed out by the authors
of \cite{hep-th/0508023} that almost all such states have an underlying
structure of the ``quantum foam'', whose universal effective description
in supergravity is a certain singular spacetime that dubs the ``hyperstar''.
They have further argued that the singularity arises because of the fact
that the classical description integrates out the microscopic details of
the quantum mechanical wavefunction.

Bubbling AdS space geometries \cite{LLM} illustrate that the
ten-dimensional type IIB metric corresponding to the half-BPS
supergravity may be described as
\begin{scriptsize}
\begin{eqnarray}
ds^2 &=& - h^{-2} (dt + V_i dx^i)^2 + h^2 (d\eta^2 + \sum_i
dx^idx^i) + \eta\ e^{G} ds_{S^2}^2 + \eta\ e^{ - G} ds^2_ {\tilde
S^3} \\ \nonumber h^{-2} &=& 2 \eta \cosh G , \\ \nonumber  \eta
\partial_\eta V_i &=& \epsilon_{ij} \partial_j z,\qquad \eta
(\partial_i V_j-\partial_j V_i) = \epsilon_{ij} \partial_\eta z
\\ \nonumber z &=&{ 1 \over 2} \tanh G
\end{eqnarray}
\end{scriptsize}
The complete solution may thus be determined by the single
function $z$, which satisfies the linear differential equation
\begin{scriptsize}
\begin{eqnarray}
\partial_i \partial_i z + \eta \partial_\eta (\frac{ \partial_\eta z}{ \eta}) =0
\end{eqnarray}
\end{scriptsize}

This is simply an electrostatic problem in $\mathrm{R}^6=
\mathrm{R}^4 \times \mathrm{R}^2$ with potential $\Phi(\eta,
x^1,x^2)= 2 \eta^{-2}$. Let the coordinates $(x^1,x^2)\in
\mathrm{R}^2$, then the smoothness of the moduli space thus
obtained may be determined from the symmetry of the background.
Here, one reveals that the appropriate boundary conditions may be
characterized by
\begin{scriptsize}
\begin{eqnarray}
z(\eta;x_1,x_2) = \frac{\eta^2}{\pi}\int dx_1' dx_2' \frac{z(0;
x_1', x_2')}{[(x_2'-x_1')^2 + \eta^2]^2}
\end{eqnarray}
\end{scriptsize}
Thus, the smoothness requires an absence of singularity and, at
$\eta =0$, it reads that $z(0; x_1,x_2)= 1/2$ for $S^3 \subset
S^5$, which in this limit shrinks to zero, while the other
admissible boundary condition comes with  $z(0; x_1,x_2)= -1/2$
for $S^3 \subset AdS_5$. The $R$-charge associated with the
underlying configuration may thus be expressed via the conformal
dimension $\Delta$ of a given state, which may be computed as
\begin{scriptsize}
\begin{eqnarray}
\Delta = \int_{ \mathcal{D}} \frac{d x^1 dx^2 }{2\pi\hbar}
\frac{1}{2}(\frac{x_1^2+x_2^2}{\hbar} - \frac{1}{2}N^2),
\end{eqnarray}
\end{scriptsize}
where $\mathcal{D}$ is the region with $\eta =0$, known as the
droplet, which satisfies $z(\eta;x_1,x_2)=0$ \cite{hep-th/0508023}.
It is also immediate that the phase-space velocity which may be
defined as $u(0,x^1,x^2)= \frac{1}{2}- z(0;x_1,x_2)$, makes the
underlying configuration smooth. Specifically, one finds that the
$u(0;x^1,x^2)$ satisfies
\begin{scriptsize}
\begin{eqnarray}
u(0; x^1, x^2) &=& 1, \ if \ \eta =1
\\ \nonumber &=& 0, \ otherwise
\end{eqnarray}
\end{scriptsize}
In order to have a proper comparison, we may consider typical
states in the boundary field theory and thus explain the matching
between the supergravity configuration and the underlying ensemble of
typical microstates. What follows next shows that the pure states with
relative energies $\{ r_i\}$ may be described by considering a
Young diagram having $N$ rows and $N_c$ columns. Such a Young
diagram may in general be depicted as
\begin{scriptsize}
\begin{eqnarray}
Y(N,N_c) &=& \Box \ \Box \ \Box \ \ \ \ldots \ \Box \\ \nonumber && \vdots \\
\nonumber && \Box \ \Box \ \Box \\ \nonumber && \Box \ \Box \\
\nonumber && \Box
\end{eqnarray}
\end{scriptsize}
Notice that the conformal dimension $\Delta \sim N^2$ puts a
constraint on the underlying ensemble, namely that, at most, it is
allowed to have $N$ number of rows. Similarly, the number of column
can be at most $N_c$, which may be treated as some sort of cutoff in
the underlying quantum theory. This is due to the fact that the
gravity singular black hole solutions may be sourced by a certain
distribution of giant gravitons. In this case, the singularity
can be described by considering spherical $D_3$-branes having
conformal dimension  $\Delta =J \sim N$, while the $1/2$
BPS condition allows for sixteen charges in the theory. Furthermore,
the existence of a RR-flux in $AdS_5$ sources certain
non-abelian interactions, which ascribes bound states of $N$ giant
gravitons. The concerned microstates of the underlying ensemble has
precisely been described in \cite{hep-th/0107119}. Subsequently,
it follows \cite{hep-th/0107119} that the number of giants on
average remains the same as the number of giants sourcing the
gravity singular solutions.

Thus, having a striking microscopic picture of the gravity singular
solutions, we shall now move on to describe the Weinhold geometric
framework for the statistical mechanics of a large number of random
Young tableaux whose (rows, columns) may respectively vary from
$(1,1)$ to $(N, N_c)$. Let us consider an ensemble of Young diagrams
in which the total number of boxes is $\Delta \in [N_c,NN_c]$. This
is because in any random Young diagram, we can have $N_c$ columns,
and thus the maximum number of admissible row would be $NN_c$, see
for instance
\begin{scriptsize}
\begin{eqnarray}
Y(N_c,1) &=& \Box \ \Box \ \Box \ \ \ \ldots \ \Box \\ \nonumber
&& 1 \ \ 2\ \ 3 \ \ \ \ldots \ \ N_c
\end{eqnarray}
\end{scriptsize}
For any other state with $\Delta= N_c+ \delta \Delta$, the system would
heat up and thus would acquire a non-vanishing effective temperature.
Consequently, the entropy would increase with $S>S_0$, and the
system would get randomized. This certainly does not happen for the
$1/2$-BPS spinless configurations, and the vacuum entropy at $T=0$
gets maximized when half of the Young diagram is full.

To be concrete, let us consider the microstates having relative
energies $\{ r_i\}_{i=1}^N$ and compute the statistical quantities
associated with the promising giant configurations. Let us define
the difference of relative fermion energies as
\begin{scriptsize}
\begin{eqnarray}
c_N &=& r_1 \\ \nonumber && \vdots \\ \nonumber c_{N-i}&=&
r_{i+1}-r_i \\ \nonumber && \vdots  \\ \nonumber c_1&=& r_N-
r_{N-1}
\end{eqnarray}
\end{scriptsize}
We thus see that the quantity $c_j$ describes the number of columns
of length $j$ in the diagram associated to $\{r_N,r_{N-1}, \ldots,
r_1\}$. As mentioned earlier, we are interested in the statistical
fluctuations of typical half-BPS states of large charge $\Delta= J
= N^2$, and thus the excitation energy of fermions is comparable
to the energy of the Fermi sea.

In order to analyze the chemical correlation in the excited states of 
a large number of free fermions, we would consider Weinhold geometry
to study the typical correlation of statistical states being
characterized by certain arbitrary Young diagrams. In order to
simplify the notations, we shall define $q= e^{-\beta}, \xi=
e^{-\lambda}, \beta = 1/T$. Then, the canonical partition
function reads
\begin{scriptsize}
\begin{eqnarray}
Z(\beta, \lambda)= \xi^{-N_c}\prod_j \frac{1}{1-\xi q}
\end{eqnarray}
\end{scriptsize}
From the very definition of the canonical ensemble
\cite{hep-th/0508023}, the average canonical energy may be
expressed by
\begin{scriptsize}
\begin{eqnarray}
<E(\beta, \lambda)>= q \partial_q \ln Z (q)= \sum_j \frac{j \xi
q^j}{1-\xi q^j}
\end{eqnarray}
\end{scriptsize}
We thus find that the average occupation number of the theory,
which measures the maximum number of allowed columns in any chosen
Young diagram, may be given to be
\begin{scriptsize}
\begin{eqnarray}
N_c= \sum_j <c_j>= \sum_j \frac{\xi q^j}{1-\xi q}
\end{eqnarray}
\end{scriptsize}
As mentioned in  the previous section, we may now divulge the
notion of chemical fluctuations without any approximation. For
this purpose, we shall redefine the canonical energy in terms
of the effective canonical temperature $T$, which leads to the
following average energy
\begin{scriptsize}
\begin{eqnarray} \label{energy}
<E(T,\lambda)>= \sum_{j=0}^{\infty} \frac{j e^{-(\lambda +
j/T)}}{1- e^{-(\lambda + j/T)}}
\end{eqnarray}
\end{scriptsize}
For investigating the chemical fluctuations, we shall consider
two neighboring statistical states characterized by the chemical
potentials $(T, \lambda)$ and $(T+ \delta T , \lambda +\delta
\lambda )$.
%
%Let us consider $q= e^{-\beta}, \chi= e^{-\lambda}, \beta = 1/T$,
%then the canonical partition function reads as, $Z(\beta,
%\lambda)= \chi^{-N_c}\prod_j \frac{1}{1-\chi q}$. The canonical
%energy $<E(\beta, \lambda)>= \sum_j \frac{j \chi q^j}{1-\chi q^j}=
%\sum_j \frac{j e^{-\lambda - j \beta}}{ e^{-\lambda - j \beta}}$.
%Redefinition of the energy in terms of thermodynamic quantity
%$T$ leads to, $<E(T,\lambda)>= \sum_{j=0}^{\infty} \frac{j
%e^{-(\lambda + j/T)}}{1- e^{-(\lambda + j/T)}}$. To investigate
%the chemical fluctuations, we shall consider two neighboring
%statistical states characterized by the chemical potentials $(T,
%\lambda)$ and $(T+ \delta T , \lambda +\delta \lambda )$.
%
The chemical pair correlation functions may thus be defined as the
components of the Weinhold metric
\begin{scriptsize}
\begin{eqnarray} \label{Wienmetric}
g_{ij}^{(E)}= \partial_i \partial_j E(T, \lambda), \ i,j= T,
\lambda
\end{eqnarray}
\end{scriptsize}
The stability of the underlying canonical chemical configuration may thus
be determined as the positivity of the determinant of the Weinhold
metric tensor, $g^{(E)}(T,\lambda) = det(g_{ij}^{(E)})$. The
fundamental configuration is just an ordinary two dimensional intrinsic
surface $(M_2,g)$ spanned by the chemical potentials $(\beta,\lambda)$.
Thus, the global correlation length may be defined as the $M_2$ scalar
curvature invariant. Explicitly, one can determine the intrinsic
covariant Riemann curvature tensor $ R_{T \lambda, T, \lambda}^{(E)}$.
Hence, the Ricci scalar may, as in Eqs. \ref{Wiencur}, be read off for
Eqs. \ref{energy} to be
\begin{scriptsize}
\begin{eqnarray}
R^{(E)}= \frac{R_{T \lambda T \lambda}^{(E)}}{g^{(E)}(T, \lambda)}
\end{eqnarray}
\end{scriptsize}
\subsection{State-space Description}
In order to have  a test of our state-space formulation, the possible
explication of the correlations associated with the box counting
entropy may be divulged as follows. Let us first consider the case
of the large $N$ limit, with $N \rightarrow \infty$, $\hbar \rightarrow 0$
and $ N\hbar$ fixed. Then, the expectation values of relative energies
$<r_i$ define a curve $<y(x)>= \sum_{i=N-x}^N <c_i>$. This essentially
leads to the evaluation of an integral $\int_{N-x}^N di <c_i>$. The
algebraic curve associated with this problem may thus be defined as
a set of energies
\begin{scriptsize}
\begin{eqnarray}
\varepsilon:= \{ \alpha q^{N-x}+\gamma q^y=1 \ | \ \alpha=
\frac{1-q^{N_c}}{1-q^{N_c+N}}, \gamma=
\frac{1-q^{N}}{1-q^{N_c+N}}\}
\end{eqnarray}
\end{scriptsize}
From the extremization of the curve $f(x,y)=  \alpha
q^{N-x}+\gamma q^y-1$, it is not difficult to see that the entropy
gets maximized when $q\rightarrow 1$. In this limit, we have a
``limit curve''
\begin{scriptsize}
\begin{eqnarray}
y(x)= \frac{N_c}{N}x= wx
\end{eqnarray}
\end{scriptsize}
This is clearly a straight line with slope $N_c/N$, and in fact it
corresponds to the triangular Young diagrams. For an illustration,
we may depict such a Young diagram for $N_c=4$ as
\begin{scriptsize}
\begin{eqnarray}
Y(4,4) &=& \Box \ \Box \ \Box \ \Box \\ \nonumber && \Box \ \Box \ \Box \\ \nonumber && \Box \ \Box \\
\nonumber && \Box
\end{eqnarray}
\end{scriptsize}
The liquid droplet model thus suggests that the involved probes
may be resolved by considering a large number of particles between
the phase-space shells $\sigma$ and $\sigma +d\sigma$. Thus the
differential curve may be expressed as
\begin{scriptsize}
\begin{eqnarray}
\frac{u(0;\sigma^2)}{2 \hbar^2}d\sigma^2= dx
\end{eqnarray}
\end{scriptsize}
Here, the droplet $u(0; x^1, x^2)$ has been expressed as $u(0;
\sigma^2)$ in terms of the radial coordinate $\sigma$, which from
the view-points of the underlying $U(1)$ symmetry presents single
Young diagram states in the $(x^1, x^2)$ plane. We thus procure an
algebraic curve satisfying previously mentioned droplet boundary
conditions with
\begin{scriptsize}
\begin{eqnarray}
\frac{\sigma^2(x)}{2 \hbar}= y(x)+x
\end{eqnarray}
\end{scriptsize}
The consistency may easily be checked and, in particular, one finds,
by differentiating the algebraic curve, that it describes the
natural boundary conditions associated with the giant black holes
\begin{scriptsize}
\begin{eqnarray}
u(0;\sigma^2)= \frac{1}{1+y'(x)}= \frac{1}{1+w}
\end{eqnarray}
\end{scriptsize}
The present analysis thus describes $1/2$-BPS $R$-charged black
holes in $AdS_5 \times S^5$, which are sometime called superstars
\cite{0109127,0411145}. It is worth to note further that the metric
of the superstar may precisely be defined by the three harmonic
functions involving the $R$-charge $q_1$ and the length $L$ of the AdS.
In this perspective, the number of allowed giants may be expressed as
\begin{scriptsize}
\begin{eqnarray}
N_c= w N; \ \forall \ w:= \frac{q_1}{L^2}
\end{eqnarray}
\end{scriptsize}
It follows \cite{0109127} that the phase-space density function
$z$ may be expressed as
\begin{scriptsize}
\begin{eqnarray}
z(\eta;x^1,x^2) = \frac{1}{2} \frac{r^2\gamma - L^2 \sin^2
\theta}{r^2\gamma + L^2 \sin ^2 \theta},
\end{eqnarray}
\end{scriptsize}
where $\gamma$ may be defined to be
\begin{scriptsize}
\begin{eqnarray}
\gamma(r)= 1+ \frac{q_1^2 \sin^2 \theta}{r^2}
\end{eqnarray}
\end{scriptsize}
For $\eta =0$, we have $r^2 \sin^2 \theta =0$, which implies that
either $r =0$ or $ \sin \theta =0$. Thus, we obtain that the
condition $ \sin \theta =0$ leads to the realization of $z(\eta;
x^1,x^2)=1/2$. This is precisely the condition that there are no
$D$-brane excitations in the ensemble of vacuum microstates. It is
thus immediate to see that the phase-space density function may
now be expressed as
\begin{scriptsize}
\begin{eqnarray}
u(0;x^1,x^2) = \frac{1}{2}- z(0;x^1,x^2)
\end{eqnarray}
\end{scriptsize}
It is worth to note for $r=0$ that the function $\gamma(r)$ leads
to the condition that the function $z(0;x^1,x^2)$ must satisfy
\begin{scriptsize}
\begin{eqnarray}
z(0;x^1,x^2)= \frac{1}{2} \frac{w-1}{w+1}
\end{eqnarray}
\end{scriptsize}
This, in turn, is consistent with the previously defined boundary
condition of the droplet $\mathcal{D}$ and provides an intriguing
microscopic picture for our present consideration. We may thus
pronounce that the Young tableaux description provides an
appropriate framework for the microscopic understanding of the
thermodynamic geometries of $1/2$ BPS supergravity configurations.

We now offer an interesting origin of the box counting entropy in
which the state-space correlations may, without any approximation, be
described. Here, an important ingredient follows from the non-trivial
notion of the coarse graining picture of the phase-space density $u_{c,g}$
of the ground state configuration, which conjointly provides an appropriate
definition of the canonical ensemble with fixed temperature, rather than
fixing the energy \cite{hep-th/0508023}. In order to be concrete, let us consider
a thin shell in the phase-space $(x^1,x^2)$ of thickness $(dx^1,dx^2)$.
Then the average density may be defined as
\begin{scriptsize}
\begin{eqnarray}
u{c,g}(0;x^1,x^2)= \frac{\int_E^{E+dE} u_{g,m}dx^2}{\int_E^{E+dE}
dx^2}
\end{eqnarray}
\end{scriptsize}
It is not difficult to ascertain that the density $u_{c,g}$ may easily
be expressed as
\begin{scriptsize}
\begin{eqnarray}
u{c,g}(0;x^1,x^2)= \frac{\Delta_X}{\Delta_E}\hbar
\end{eqnarray}
\end{scriptsize}
We may however re-express the phase-space density $u_{c,g}$ and
appreciate that it satisfies the familiar form
\begin{scriptsize}
\begin{eqnarray}
u{c,g}(0;x^1,x^2)= \frac{1}{1+y'(x)},
\end{eqnarray}
\end{scriptsize}
where $y(x)= \frac{\sigma^2}{2\hbar^2}-x$ is the aforementioned
relative energy curve for the $1/2$-BPS giant configurations at
very large $N$. The origin of the entropy may thus be ascribed as
follows. Let us consider the phase-space density and choose a set of coherent
basis which picks up the least uncertainty, and thus turns out to be
appropriate for the present discussion, following from the coarse
graining phenomenon. In order to divulge the intrinsic geometric notion of
degeneracy, we may focus our attention on an arbitrary phase-space,
having $M^2$ cells with at most random $n$ excited cells. The
pictorial view of this may be given as follows:
\begin{scriptsize}
\begin{eqnarray}
Y(N,N_c) &=& \Box \ {\color{blue} \maltese} \ \Box  \ \ldots \\
\nonumber && {\color{blue} \maltese} \ \Box \ {\color{blue}
\maltese} \\ \nonumber && \Box \ \Box  \ \Box \\ \nonumber && \ \
\ \ \ \ \ \ \ \ \ddots \\ \nonumber && \ \ \vdots \ \ \ \  \\
\nonumber && \downarrow_{(RG \ trans.)} \\ \nonumber && \downarrow
\\ \nonumber && \Box \ \Box \ \Box \ \ldots  \\ \nonumber && \Box
\ \Box \ \Box \\ \nonumber && \Box \ \Box \ \Box \\ \nonumber && \
\ \ \ \ \ \ \ \ \ddots \\ \nonumber && \ \vdots \ \ \ \  \\
\nonumber &=& Phase \ Space \ (N,N_c)
\end{eqnarray}
\end{scriptsize}
In an arbitrary Young diagram $Y(N,N_c)$, we thus see that there
are $n$ boxes filled with maltese and the rest $(M^2-n)$ of them are
empty boxes. Therefore, the existing degeneracy in choosing random
$n$ maltese (excited boxes) out of the total $M^2$ boxes may  be
given by
\begin{scriptsize}
\begin{eqnarray}
\Omega(n,M)= \ ^{M^2}C_n = \frac{(M^2)! }{(n)! (M^2-n)! }
\end{eqnarray}
\end{scriptsize}
From the first principle of statistical mechanics, we thus find
that the canonical counting entropy may be defined to be
\begin{scriptsize}
\begin{eqnarray} \label{entropy}
S(n,M)= \ln (M^2)! - \ln (n)! - \ln (M^2-n)!
\end{eqnarray}
\end{scriptsize}
It is nevertheless important to emphasize that the subsequent analysis
does not exploit any approximation, such as Stirling's approximation or
the thermodynamic limit, and thus we shall work in the full picture of the
canonical ensemble. Consequently, the present analysis bequeaths an
exact expression for the state-space pair correlation functions and
correlation length.

It is now immediate to divulge an appropriate appraisal for the
state-space geometry associated with the random Young tableaux
$Y(N, N_c)$ thus described with arbitrary $n$ excited droplets
among the $M^2$ fundamental cells in an ensemble. We observe that
the state-space interactions allow us to analyze the correlation
between two neighboring statistical states $(n,M)$ and $(n+ \delta n,
M+ \delta M)$, and thus the fluctuations in the droplets with $(n,M)$
may be defined via the well-defined Ruppenier metric as
\begin{scriptsize}
\begin{eqnarray}\label{Ruppmetric}
g_{ij}^{(S)}= - \partial_i \partial_j S(n,M), \ i,j= n, M
\end{eqnarray}
\end{scriptsize}
As we have outlined in the case of chemical fluctuations from the
Weinhold geometry, a very similar analysis may thus easily be
performed to reveal the state-space correlations existing among
the fluctuating microstates of the $1/2$ BPS configuration
describing certain giants or the superstars.

The conformal relation of the two geometries may formally be
derived via the temperature as a conformal factor, which may be
accomplished by taking the derivative of the entropy with respect
to the energy $ \Delta $. One may however easily expose that
$q= exp(-\frac{\alpha}{N})$, and it comes as no surprise that this yields
\begin{scriptsize}
\begin{eqnarray} \label{effTemp}
T^{-1}= \frac{\partial S}{\partial \Delta}= \frac{ \alpha}{N}
\end{eqnarray}
\end{scriptsize}
Nevertheless, it is important to note that there is no physical
temperature involved with half-BPS states. However, $T$ should be
understood as an effective temperature of a canonical ensemble
that sums up the half-BPS states with the canonical energy sharply
peaked at $\Delta $. More generally, if $\Delta $ scales as $N^{\xi}$,
then the Ref. \cite{hep-th/0508023} shows that the effective temperature
grows as $N^{\xi /2}$.

Our thermodynamic geometries, thus explained in either the
description of chemical potentials or the number of boxes in an
arbitrarily excited Young tableau, entail remarkably simple
expressions. Furthermore, we discover that there exists an
appropriate microscopic/macroscopic explanation for the
statistical correlations of the $1/2$ BPS supergravity
configurations. From the viewpoint of $AdS_2/CFT_1$,
it could be interesting to explore the case of more charged
black holes, e.g. extreme Reissner-Nordstr\"om black hole
\cite{SB2}, non-BPS supergravity black holes \cite{07104967v3}
or a further addition of electric-magnetic charges. 
In the next section, we shall intimately discuss the above
outlined study for the two parameter giants and superstar
solutions. In particular, our investigation shows that there
exists an exact account for the thermodynamic correlations of
the spinless black holes arising from the coarse grained
LLM geometries \cite{LLM}.

%An appropriate consideration of the state-space geometry may be
%divulged by the random Young tableaux with certain arbitrary
%excited boxes. In order to simplify the idea, let us consider a
%Young tableau involving $M^2$ fundamental cells, then the
%degeneracy in choosing $n$ excited cells is given by $ \Omega
%(n,M)= \ ^{M^2}C_n = \frac{(M^2)!}{(n)! (M^2-n)!}$. Thus the
%statistical entropy concerned with counting $n$ excited boxes is
%defined to be $S(n,M)= \ln \Omega(n,M)= \ln (M^2)! - \ln (n)!- \ln
%(M^2n)!$. In order to analyze the correlation between two
%neighboring statistical states $(n,M)$ and $(n+ \delta n , M+
%\delta M)$, and thus the fluctuations in $(n,M)$ may be defined
%via the well-defined Ruppenier metric as, $g_{ij}^{(S)}= -
%\partial_i \partial_j S(n,M) \ i,j= n, M $. A similar analysis may
%thus easily be performed as we outlined in the case of chemical
%fluctuations.
%
\section{Canonical Energy Fluctuations}
In this section, we shall present the essential features of
the chemical geometry and thereby put them into effect for the 2-
parameter giants and superstars. Here, we focus our attention
on the geometric nature of a large number of gravitons, within the
neighborhood of small fluctuations with given chemical potentials
introduced in the framework of $1/2$-BPS configurations. As stated
earlier, the thermodynamic metric in the chemical potential space is
given by the Hessian matrix of the canonical energy, with respect to
the intensive variables, which in this case are the two distinct chemical
potentials carried by the giants and superstars. We find, in this framework,
that the exact canonical energy as the function of chemical potentials
takes an intriguing expression
\begin{scriptsize}
\begin{equation}
E(T, \lambda)= \sum_{j=0}^{\infty} \frac{j \ exp(-( \lambda
+j/T))}{1-exp(-( \lambda +j/T))}
\end{equation}
\end{scriptsize}
For to obtain the thermodynamic metric tensor in the chemical
potential space, we may employ the formula in Eq. \ref{Wienmetric},
which leads to the following series expression for the components
of the metric tensor:
\begin{scriptsize}
\begin{eqnarray}
g_{TT}(T, \lambda)&=&\sum_{j=0}^{\infty}\big( -2 \frac{j^2}{T^3}
\frac{exp(- \lambda-j/T)}{(1-exp(- \lambda-j/T))}+ \frac{j^3}{T^4}
\frac{exp(- \lambda-j/T)}{(1-exp(- \lambda-j/T))}\\ \nonumber && +
3\frac{j^3}{T^4}\frac{exp(- \lambda-j/T)^2}{(1-exp(-
\lambda-j/T))^2}
+2\frac{j^3}{T^4}\frac{exp(- \lambda-j/T)^3}{(1-exp(- \lambda-j/T))^3}\\
\nonumber && -2\frac{j^2}{T^3}\frac{exp(- \lambda-j/T)^2}
{(1-exp(- \lambda-j/T))^2}\big)
\end{eqnarray}
\end{scriptsize}
\begin{scriptsize}
\begin{eqnarray}
g_{T \lambda}(T, \lambda)&=& \sum_{j=0}^{\infty}
\big(-\frac{j^2}{T^2} \frac{exp(- \lambda-j/T)}{(1-exp(-
\lambda-j/T))}-3
\frac{j^2}{T^2} \frac{exp(- \lambda-j/T)^2}{(1-exp(- \lambda-j/T))^2} \\
\nonumber && -2\frac{j^2}{T^2}\frac{exp(- \lambda-j/T)^3}
{(1-exp(- \lambda-j/T))^3}\big)
\end{eqnarray}
\end{scriptsize}
\begin{scriptsize}
\begin{eqnarray}
g_{ \lambda \lambda}(T, \lambda)&=& \sum_{j=0}^{\infty} \big( j
\frac{exp(- \lambda-j/T)}{(1-exp(- \lambda-j/T)}
+3 j \frac{exp(- \lambda-j/T)^2} {(1-exp(- \lambda-j/T))^2} \\
\nonumber && +2 j \frac{exp(- \lambda-j/T)^3} {(1-exp(-
\lambda-j/T))^3} \big)
\end{eqnarray}
\end{scriptsize}
Furthermore, it turns out that the determinant of the metric
tensor takes an inelegant polynomial form and thus, in order
to simplify the notations, we may define a level function
\begin{scriptsize}
\begin{eqnarray}
b_j(T, \lambda):= exp(- \lambda-j/T)
\end{eqnarray}
\end{scriptsize}
It is however evident that the local stability of the full
phase-space configurations may be determined by computing the
determinant of the concerned thermodynamic metric tensor. Here,
we may likewise provide a compact formula for the determinant of
the metric tensor, and exclusively, the intrinsic geometric analysis
assigns a compact expression to the determinant of the metric tensor.
A straightforward computation thus shows that the determinant of the
metric tensor as the function of arbitrary chemical potentials is
\begin{scriptsize}
\begin{eqnarray}
||g(T, \lambda)|| &=& \sum_{j=0}^{\infty}(-j^2 \frac{b_j}{T^4}
\frac{(-2 T+2 T b_j+j+ jb_j)}{(b_j-1)^3)}) \times \\ \nonumber &&
\sum_{j=0}^{\infty}(-jb_j \frac{(1+b_j)}{(b_j-1)^3})-
(\sum_{j=0}^{\infty}(j^2 \frac{b_j}{T^2}
\frac{(1+b_j)}{(b_j-1)^3)})^2
\end{eqnarray}
\end{scriptsize}
\begin{figure}[htb]
\vspace*{-1cm}
\begin{center}
\hspace{-1.0cm}
%\vspace*{1cm}
\epsfig{file=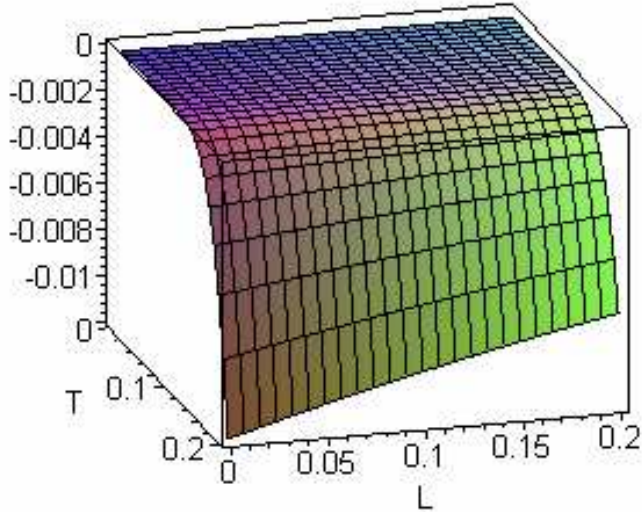,width=12cm}
\end{center}
\vspace*{0.01cm}
\caption{The Wienhold determinant of the metric tensor as a function of temperature,
T; and chemical potential, $ \lambda=L$ in the $1/2$-BPS supergravity configurations.}
\vspace*{0.50cm}
\end{figure}
We thus see from Fig. 1 that the determinant of the Weinhold metric develops
an instability when $T>0.15$. This is also expected, since the coarse graining
phenomenon should break down near small effective canonical temperatures. The
system is well-defined and physically sound for all $ T \in [0, 0.15]$.

In order to conclusively analyze the nature of chemical correlations
and the concerned properties of the statistical configuration, one needs
to determine certain globally invariant quantities on the intrinsic manifold
$(M_2,g)$ of the  parameters. In fact, one may easily ascertain that
the simplest of such invariants is the scalar curvature of $(M_2,g)$. In order to
do so, one may first compute the Riemann curvature tensor $R_{T \lambda T
\lambda}(T,\lambda)$ and then, from the previously defined expression, obtain
the scalar curvature. We see that it is not difficult to compute the
covariant Riemann tensor $R_{T\lambda T \lambda}(T, \lambda)$. The
exact expression for the $R_{T\lambda T \lambda} (T, \lambda)$ is quite
involved and thus we relegate it to the Appendix B.

In a straightforward fashion we can show, by applying our previously advertised
intrinsic geometric technology, that one may easily obtain the scalar
curvature simply via the relation
\begin{scriptsize}
\begin{eqnarray}
R(T, \lambda)= \frac{R_{T  \lambda T \lambda}(T, \lambda)}{||g(T,
\lambda)||}
\end{eqnarray}
\end{scriptsize}
\begin{figure}[htb]
\vspace*{0cm}
\begin{center}
\hspace{-1.0cm}
%\vspace*{1cm}
\epsfig{file=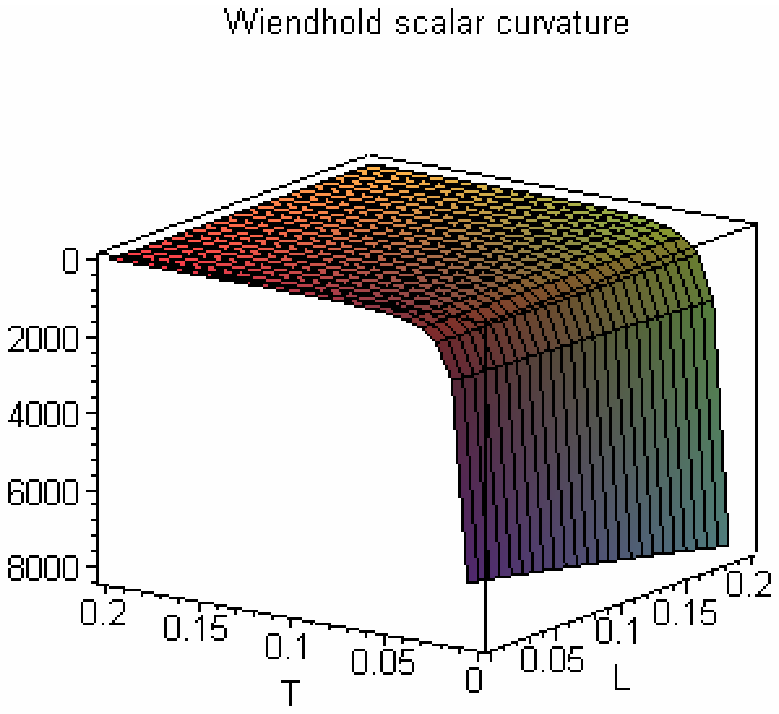,width=12cm}
\end{center}
\vspace*{0.01cm}
\caption{The Wienhold scalar curvature as a function of temperature, T; and
chemical potential, $ \lambda=L$ in the $1/2$-BPS supergravity configurations.}
\vspace*{0.50cm}
\end{figure}
We observe from Fig. 2 that the system becomes strongly correlated
for $T<0.05$, and for $ T \in [0.01, 0.02]$ it acquires the chemical
correlation length of $8 \times 10^3$. Interestingly, we do not see
any chemical fluctuations for $T>0.08$.
In the domain of small chemical potentials $T, \lambda \in [0,0.2] $,
the pictorial views of the stability may be perceived from the
determinant of the Weinhold metric tensor $g^{(E)}(T, \lambda)$
and the concerned correlation length from the corresponding scalar
curvature $R^{(E)}(T, \lambda)$. We notice that both the plot of
the determinant and that of the scalar curvature of the Wienhold 
geometry are nice regular 3D hyper-surfaces.

We thence realize that the microcanonical representation of the black
holes under consideration posits that there exists a large number
of correlated degenerate microscopic states of the $1/2$ BPS giants
and superstars. The non-vanishing scalar curvature suggests that
a constituent microstate may as well avoid that the initial pure
states collapses to a particular pure black hole microstate. This
is because of the fact that the intrinsic geometric structures of
AdS black hole can be deduced by certain suitably defined subtle
measurements \cite{hep-th/0508023}, and thus there is no loss of
information.
\section{The Fluctuating Young Tableaux}
As in the previous section, the present section is devoted to investigate
the notion of large number of boxes into our state-space geometry, and thereby
we shall analyze the possible role of concerned large integers in the veritable
giant black hole configurations, being considered in the framework of liquid
droplet model or fuzzball description. In this concern, we shall precisely
compute the statistical pair correlations under this consideration, which
ascribes a set of self convincing physical meaning to the state-space
quantities, for the simplest excited giant configurations
\begin{scriptsize}
\begin{eqnarray}
S(n,M)= ln((M^2)!)-ln(n!)-ln((M^2-n)!)
\end{eqnarray}
\end{scriptsize}
Employing the previously proclaimed formulation, one may easily read
off the components of the covariant state-space metric tensor to be
\begin{scriptsize}
\begin{eqnarray}
g_{nn}(n,M)= \Psi(1,n+1)+ \Psi(1,M^2-n+1)
\end{eqnarray}
\end{scriptsize}
\begin{scriptsize}
\begin{eqnarray}
g_{nM}(n,M)=  -2 M \Psi(1,M^2-n+1)
\end{eqnarray}
\end{scriptsize}
\begin{scriptsize}
\begin{eqnarray}
g_{MM}(n,M)= 4 M^2 \Psi(1,M^2-n+1) -4 M^2 \Psi(1,M^2+1)-2
\Psi(M^2+1)+2 \Psi(M^2-n+1)
\end{eqnarray}
\end{scriptsize}
In the above expressions, the $\Psi(n,x)$ is the $n^{th}$
polygamma function, which is the $n^{th}$ derivative of the usual
digamma function. Specifically, it turns out that the $\Psi(x)$
may defined to be
\begin{scriptsize}
\begin{eqnarray}
\Psi(x)= \frac{\partial }{ \partial x} \ln( \Gamma(x))
\end{eqnarray}
\end{scriptsize}
In this framework, we observe that there exists a very simple
description which divulges the geometric nature of the statistical
pair correlations. The fluctuating extremal black holes may thus
easily be determined in terms of the mass and angular momentum of
the underlying configurations. Moreover, it is evident that the
principle components of the statistical pair correlations are
positive definite, for a range of the parameters of the concerned black
holes, which physically signifies a certain self-interaction of a
fictitious particle moving on an intrinsic surface $(M_2(R),g)$.
Significantly, it is clear in this case that the following
state-space metric constraints hold:
\begin{scriptsize}
\begin{eqnarray}
g_{nn} &>& 0, \ \ \forall \ (n,M)  \ |  \ \Psi(1,n+1)+ \Psi(1,M^2-n+1)>0  \\
\nonumber  g_{MM} &>& 0, \ \ \forall \ (n,M)  \ |  \
\Psi(1,M^2-n+1)-\Psi(1,M^2+1)> \\ \nonumber &&
\frac{1}{2M^2}(\Psi(M^2+1)- \Psi(M^2-n+1))
\end{eqnarray}
\end{scriptsize}
Consequently, we may easily reveal that the common domain of the above
state-space constraints defines the range of physically sensible
values of the chosen number of boxes and total number of boxes, such
that the giants may remain into certain locally stable statistical
configurations. We may also notice that the total boxes component
$g_{MM}$ of the state-space metric tensor is asymmetrical, in comparison
to the $g_{nM}$ and $g_{nn}$. This is physically well-accepted, because
of the fact that the component associated with the large number of boxes
is somewhat like the heavy head-on collision of two equal particles, which
alternate more energy, in contrast to the other excitations, either involving
the single particle or the excited-unexcited particles. It is pertinent to
mention that the relative pair correlation function determines the selection
parameter for a chosen black hole, which may be defined as the modulus of the
ratio of excited-excited to excited-unexcited statistical correlations.
Importantly, we procure that the parameter thus apprised may be given
as the ratio of two digamma functions
\begin{scriptsize}
\begin{eqnarray}
a: = \frac{1}{2M} \vert \frac{\Psi(1,n+1)}{\Psi(1,M^2-n+1)}\vert
\end{eqnarray}
\end{scriptsize}
It is worth to mention that when $n>1$, one has $\Psi(n,x)= \Psi(
n ) + \gamma$, which is an ordinary rational number, where the
$\gamma$ is the standard Euler's constant. For small values of $n$,
$\Psi(n)$ is computed as a sum of gamma functions, which is again a rational
number. To force this computation to be performed for larger values
of the $n$, we may use
\begin{scriptsize}
\begin{eqnarray}
\Psi(n,x) = \frac{\partial^n \Psi(x)}{\partial x^n},
\end{eqnarray}
\end{scriptsize}
with $\Psi(0,x) = \Psi(x)$. Moreover, the stability of the
underlying statistical configurations may earnestly be analyzed by
computing the degeneracy of the associated two dimensional state-space
manifold. In fact, we may easily ascertain that the determinant of
the state-space metric tensor may be given to be
\begin{scriptsize}
\begin{eqnarray}
g(n,M) &=& -4M^2 \Psi(1,n+1)\Psi(1,M^2+1)-2\Psi(1,n+1)\Psi(M^2+1) \\
\nonumber && +4M^2 \Psi(1,n+1)\Psi(1,M^2-n+1)
+2\Psi(1,n+1)\Psi(M^2-n+1) \\ \nonumber &&
-4M^2 \Psi(1,M^2-n+1)\Psi(1,M^2+1) -2\Psi(1,M^2-n+1)\Psi(M^2+1) \\
\nonumber && +2\Psi(1,M^2-n+1)\Psi(M^2-n+1)
\end{eqnarray}
\end{scriptsize}
\begin{center}
 \begin{figure}[htb]
\vspace*{-3.5cm}
\begin{center}
\hspace{-7.5cm}
%\vspace*{1cm}
\epsfig{file=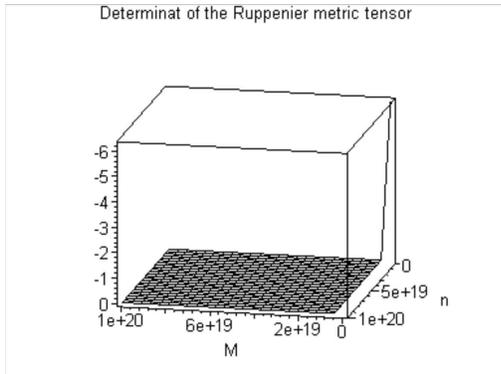,width=14cm,angle= 90}
\end{center}
\vspace*{-4cm}
\caption{The Ruppenier determinant of the state-space metric tensor
as a function of excited droplets, $n$; and total number of fundamental
cells, $M^2$ in the $1/2$-BPS supergravity configurations.}
\vspace*{0.50cm}
\end{figure}
\end{center}
Fig. 3 indicates that the canonical configurations become unstable
and ill-defined when both of the $n$ and $M$ take small values. Up
to $10^5$ boxes, we discover that the same qualitative features
hold. Particularly, it is worth to mention that the fluctuation
bumps are aligned towards the $n^{th}$ boundary, when $n$ and $M$
are of the same order.

The determinant of the metric tensor thus calculated is non-zero
for any set of given non-zero total boxes and excited boxes, and
thus, for the set of proper choices, it provides a non-degenerate
state-space geometry for this configuration. In turn, one may
illustrate the order of statistical correlations between the
equilibrium microstates of the giant black hole system. Besides
the fact that the principle component constraints $\lbrace g_{ii}
> \ 0 \ \vert \ i= n, M \rbrace$ imply that this system may
accomplish certain locally stable statistical configurations,
however the negativity of the determinant of the state-space
metric tensor indicates that the underlying systems may globally
endure a certain instability, as well for the bad choice of boxes,
which correspond to certain unphysical macroscopic configurations.

This significantly connotes that there exists a positive definite
volume form on the $(M_2,g)$, for the good choice of boxes and
their excitations. One may thus conclude that this system may
remain in the nice chosen configuration or, for certain
unacceptable choices, might move to some more stable brane
configurations. Furthermore, one may easily examine, in general,
that the covariant Riemann correlation tensor may be given by the
two large integers characterizing the droplets. However, an
explicit expression is given by Eq. \ref{countentcovcur} for the
$R_{nMnM}(n,M)$, which is rather involved, and thus we have
relegated it to the Appendix B.

Furthermore, in order to examine certain global properties of
such black holes phase-space configurations, one is required to
determine the associated geometric invariants of the underlying
state-space manifold. For the giant and superstar black holes,
the simplest invariant turns out to be the state-space scalar
curvature, which may easily be computed by using the intrinsic
geometric technology defined as the negative Hessian matrix of the
entropy captured by the excited contributions. Indeed, we discover
that the state-space curvature scalar for the fluctuating giant
(and superstar) configurations may easily be depicted.
Nevertheless, the exact involved expression for the scalar
curvature $R(n,M)$ has been depicted in Eq. \ref{countentscacur}
and is shifted to the Appendix B.
\begin{figure}[htb]
\vspace*{.30cm}
\begin{center}
\hspace{-1.0cm}
%\vspace*{1cm}
\epsfig{file=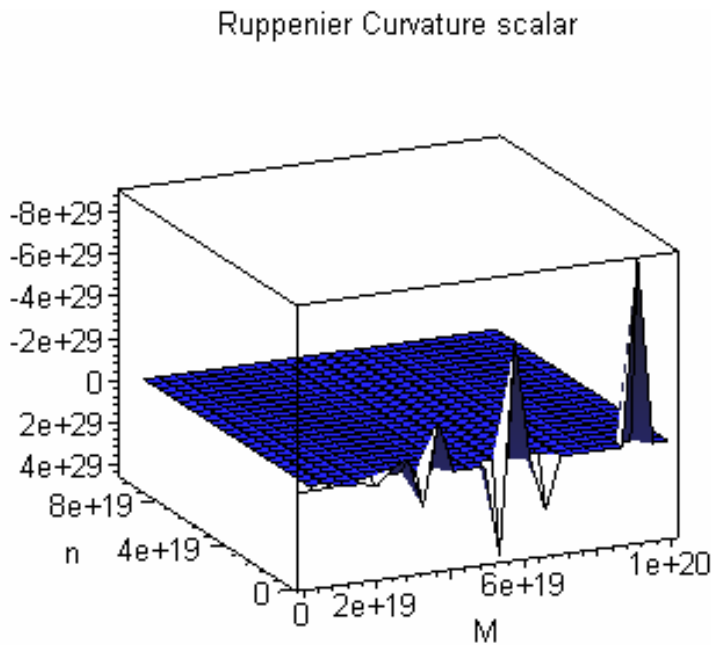,width=12cm}
\end{center}
\vspace*{-0.1cm} \caption{The Ruppenier scalar curvature as a
function of excited droplets, $n$, and the total number of
fundamental cells, $M^2$, in the $1/2$-BPS supergravity
configurations.} \vspace*{0.50cm}
\end{figure}
We numerically find the following conclusions for the state-space
scalar curvature correlations.

\begin{itemize}
\item For $n, M \in [1, 10^{10}]$, it turns out from Fig. 4 that
positive, as well as negative correlations exist only along the
$M^{th}$ boundary of the state-space configuration. It is also
expected from the general consideration of statistical fluctuation
theory that the system would possess certain attractions and
repulsions.
\item We notice, from the respective plots of the determinant and scalar
curvature of the state-space configuration, that the system
remains stable and regular along the $n^{th}$ boundary of the
state-space configuration. We need not mention in this range that
the system becomes ill-defined near the origin $(0, 0)$ of the
state-space manifold.
\item We further observe for $10^4$ boxes that there exist definite small
attraction and repulsion along the $n^{th}$ boundary of the $R$-$n$
surface. Interestingly, the same conclusion remains true for the $g$-$n$
surface of the underlying state-space configuration.
\end{itemize}
We recognize, in the framework of our state-space geometry, that
the negative sign of the curvature scalar signifies that this
system is effectively an attractive configuration, under the
Gaussian fluctuations. Thus, the giants and superstars appear to
be manifestly stable configurations, with nice combinatorial
properties of the Young tableaux. Furthermore, we observe that the
curvature scalar thus considered is inverse squarely proportional
to the determinant of the underlying state-space metric tensor,
and thus there are no genuine state-space instability in the
underlying microscopic configurations. In turn, we discover that
the underlying statistical correlations remain definite, non-zero,
finite, regular functions of total boxes and excited boxes carried
by the giant black holes.

It is important to note that the state-space geometric quantities
may become ill-defined, if the state-space co-ordinates, being
defined as the space-time parameters, jump from one existing
domain to another domain of the solution. This indicates that the
microscopic configurations may correspond to certain interacting
statistical systems, in the chosen branch of the full black hole
solutions. In an admissible domain of physically acceptable boxes,
we discover that the giant black holes have no phase transition,
and thus the fundamental statistical configurations are completely
free from critical phenomena. Note that the absence of divergences
in the scalar curvature indicates that the BPS black brane
solutions endure an everywhere thermodynamically stable systems,
on their respective state-space configurations.

More generally, we find that the regular state-space scalar
curvature seems to be comprehensively universal for the given
number of parameters of the configuration. In fact, the concerned
idea turns out to be related with the typical form of the
state-space geometry, arising from the negative Hessian matrix of
the duality invariant expression of the black hole entropy. As in
the standard interpretation, the state-space scalar curvature
describes the nature of underlying statistical interactions of the
possible microscopic configurations, which characteristically turn
out to be non-zero, for the $1/2$-BPS black holes. In fact, we may
easily appreciate that the constant entropy curve is a standard
curve, which may be given by
\begin{eqnarray}
(M^2)!= c\ n!(M^2-n)!
\end{eqnarray}
Here, the real constant $c$ can be determined from the given
expression of the vacuum entropy. This determines the giant black
hole embedding in the view-points of the state-space geometry.
Moreover, we may also disclose in the present case that the curve
of constant scalar curvature is, however, a complicated curve on
the space of total boxes and excited boxes, but the the concerned
fundamental nature may easily be fixed by the given number of the
boxes. Nonetheless, it is not difficult to enunciate the
quantization condition existing on the charges, as we have an
integer number of boxes, which in turn signifies a general
coordinate transformation in the large number of boxes on the
state-space manifold, and may thus be presented in terms of the
net number of respective branes.
\section{Remarks and Conclusion}
This paper provides an exact intrinsic thermodynamic geometric
account of the canonical energy fluctuations and box counting
entropy for the giants and superstars.  The arguments required
have been accomplished from an ensemble of Young tableaux with
certain randomly excited boxes, while the remaining boxes were
kept intact. We notice that the microcanonical representations of
the black holes under consideration posit a large number of
correlated degenerate microscopic states. These microstates may be
evaded by showing that an initial pure state collapses to a
particular pure black hole microstate, whose exact structure can
be deduced by suitably subtle measurements \cite{hep-th/0508023},
along with the loss of information, if any. Here, our focus has
been to investigate the implications of correlated states for the
$1/2$-BPS supergravity black holes. Our analysis clarifies the
nature of underlying equilibrium configurations over the Gaussian
fluctuations.

Given the importance of the canonical correlations, we have
addressed the following questions: (i) how do the Gaussian
correlations of pure microstates look like in the view-points of
thermodynamic geometries, (ii) what sorts of correlation areas may
distinguish the giants or superstars from each other, and (iii)
what is the typical nature of statistical fluctuations over an
equilibrium ensemble of CFT microstates? Interestingly, we have
explicitly demonstrated that the underlying chemical fluctuations
involve ordinary summations, while the state-space fluctuations
may simply be depicted by standard polygamma functions. Hereby, we
notice that the large black holes in AdS spacetimes, whose horizon
sizes are bigger than the scale of the AdS curvature, are stable,
and exclusively there is no thermodynamic phase transition. We
herewith find a precise matching with the fact that these black
holes come into an equilibrium configuration, because the AdS
geometries create an effective confining potential \cite{HP}, and
thus yield the stability of the thermodynamic system.

Importantly, our framework exploits the fact that the classical
general relativity large black holes come with the basic property
that their horizon area satisfies the Bekenstein-Hawking entropy
relation. With an understanding of the limiting microscopic
configurations, an appropriate intrinsic geometric notion has in
effect been offered to the counting problem of the microstates of
the giants and superstars. Following the discussion of the Appendix
A, this provides an account to the degeneracy of microscopic states
in the limit $G_N/l_P>>1$. The semi-classical approximation thus 
uncovers that the Boltzmann statistical entropy entails certain 
well-defined equilibrium microstate systems, upon the inclusion
of a quadratic fluctuation on the phase-space configurations.
Our analysis thus shows that the $1/2$-BPS giants and superstars
are thermodynamically stable objects and do not emit Hawking radiation.

The state-space description shows that a set of exact expressions
may easily be given for the quadratic fluctuations, over an
equilibrium canonical configuration characterized by possible
excited and unexcited droplets, in an ensemble of arbitrary random
Young tableaux. In turn, the state-space fluctuations, stability
criterion and state-space correlation length have precisely been
determined, without any approximation, for a large number of giant
gravitons and superstars. Following the standard Riemannian
geometric notions, we expose that the chemical configuration
yields in general that the chemical pair correlation functions,
stability condition and correlation length, for an arbitrary value
of the effective canonical temperature and the chemical potential,
may be determined over an intrinsic Weinhold manifold. Our
intrinsic geometric study, thus, exemplifies that there exists an
exact fluctuating statistical configuration, which involves an
ensemble of fuzzballs, or a number of liquid droplets.

Furthermore, we can ask how the underlying highly degenerate
equilibrium microstates get correlated with the chemical
potential, carried by a set of given excited droplets. We have
shown, from the perspective of string theory on $AdS_5 \times
S^5$, that the five dimensional $R$ charged AdS black holes admit
an exact Legendre transformed dual chemical configuration, which
may be described in terms of the  chemical potential and an
effective canonical temperature. Thus, the origin of gravitational
thermodynamics comes with the existence of a non-zero
thermodynamic curvature, under the coarse graining mechanism of
alike ``quantum information geometry'', associated with the wave
functions of underlying BPS black holes. Notice, further, that the
physical meaning of the respective curvatures is that they
describe the correlation, in the concerned ensemble of black
holes. Characteristically, it is worth to mention that the
semi-classical gravitational description, arising with the
non-vanishing scalar curvature, signifies the existence of the
throat of the underlying AdS background.

We notice an intriguing support, from the Mathur's fuzzball
proposal, that the leading order entropy should come from those
fuzzs, whose radius varies as the fuzzy throat of the black hole
horizon size. Furthermore, the liquid droplet model suggests that
the AdS length scales like $L\sim l_P f(N_i)$, where $N_i$ is the
number of bound states \cite{hep-th/0508023, hep-th/0107119}. It
has further been suggested that the classical description is
achieved, when $N \rightarrow \infty$ while $\hbar \rightarrow 0$,
such that the Fermi level $N/ \hbar$ remains constant. Most
importantly, we have investigated the role of thermodynamic
fluctuations in the two parameter giants and superstars, whose
chemical and state-space configurations are being respectively
characterized by the chemical variables of the effective canonical
ensemble, and by the number of both excited and total boxes,
constituting an ensemble of arbitrary shaped CFT microstates. The
present analysis thus explicates that the underlying chemical and
state-space configurations are well-defined, non-degenerate and
regular, for all physically admissible domains of the statistical
parameters, defining an ensemble of arbitrary fuzzballs, or a set
of indiscriminate liquid droplets.

Now, we enlist a number of attributes arising from our study of
the thermodynamic intrinsic geometry of (dual) giants and
superstars, divulged in the framework of liquid droplets or
fuzzball solutions. The intriguing nature of the chemical and the
state-space correlations, existing in an underlying statistical
basis, ascribes that the local and global thermodynamic
structures, thus revealed, may in either case be summarized as
follows.
\subsection{Chemical Description:}
Following our specific notions, we exclusively observe that the
chemical configurations of the $1/2$ BPS giants and superstars
realize the following general properties:
\begin{itemize}
\item There exists an exact expression for the chemical pair
correlation functions, stability condition and correlation length,
for arbitrary values of $T, \lambda \in M_2$.
\item Explicit plots displayed in the Figs. 1 and 2 show that
the determinant and scalar curvature are non-trivially curved, and
surprisingly the results remain the same, even for a single
component with $j=1$.
\item We further reveal, from the numerical conclusions displayed in
Figs. 1 and 2, that the Weinhold metric tensor, and the
corresponding correlation length, show that the statements of
stability and regularity hold for all $\lambda \in M_2$.
\end{itemize}
\subsection{State-space Description:}
In this case, the findings obtained from the intrinsic state-space
geometry from an ensemble of Young tableaux may be summarized as
\begin{itemize}
\item A set of exact precise expressions may easily be given for the
state-space fluctuations, over an equilibrium canonical ensemble,
characterized by possible arbitrary random Young tableaux.
\item The state-space fluctuations, stability criterion and
state-space correlation length may easily be determined without
any approximation.
\item We can express the $g^{(S)}_{ij}(n,M)$, $g^{(S)}(n,M)$
and $R^{(S)}(n,M)$ in terms of nice, well-behaved digamma
$\Psi(n)$ and poly-gamma $\Psi(n,M)$ functions. Notice that the
mixing between excited and non-excited droplets or fuzzs may
precisely be caused by these standard polygamma functions.
\item The concerned state-space geometry corresponds to a non-degenerate,
locally stable and attractive statistical configuration.
\item The underlying numerical computations plotted in Figs. 3 and 4
disclose that the state-space correlation exists along the
boundary of $n$ and $M$, for non-large $n$, $M$, when we take
approximately $1$ to $10^5$ boxes in the Young diagrams.
\end{itemize}
It is worth to mention that the present analysis takes into
account the scales, that are larger than the respective
correlation length, and contemplates that just a few giant or
superstar microstates cannot dominate the entire macroscopic
solution. Most importantly, we have procured that the
thermodynamic intrinsic geometric structures of the canonical
energy and box counting entropy provide a coherent framework, to
further study the thermodynamic geometries, arising from the large
number of microstates of the chosen giants, superstars and the
other superconformal field theory and supergravity configurations
\cite{9711200v3}.

Finally, it is worth to mention that the interpretation of a
non-zero intrinsic scalar curvature, with an underlying
interacting statistical system, remains valid even for higher
dimensional intrinsic Riemannian manifolds. The implication of a
divergent intrinsic covariant curvature may accordingly be
divulged, from the Hessian matrix of the canonical energy or the
box counting entropy, irrespectively whether, or not, there exists
a phase transition in the $1/2$ BPS black hole configurations.
\section*{Acknowledgements}
This work has been supported in part by the European Research
Council grant n.~226455, ``SUPERSYMMETRY, QUANTUM GRAVITY AND
GAUGE FIELDS (SUPERFIELDS)''. We would like to thank Prof. J.
Sim\'on for useful discussions and view-points provided during the
``School on Attractor Mechanism SAM-2009, INFN- Laboratori
Nazionali di Frascati, Roma, Italy''. BNT would like to thank
Prof. S. Mathur and Prof. A. Sen for useful discussions offered
during the ``Indian String Meeting, ISM-2006, Puri, India''; Prof.
J. de Boer during the ``Spring School on Superstring Theory and
Related Topics-2007 and 2008, ICTP, Trieste, Italy''; and Yogesh
Srivastava during the Indian String Meeting-2007, Harish-Chandra
Research Institute, Allahabad, India; and Mohd. A. Bhat and V.
Chandra for reading the manuscript and making interesting
suggestions. BNT especially thanks Prof. V. Ravishankar for
encouragements and necessary supports provided during this work.
BNT would like to acknowledge nice hospitality of the
``INFN-Laboratori Nazionali di Frascati, Roma, Italy'' being
offered during the ``School on Attractor Mechanism: SAM-2009''
where part of this work was performed. The research of BNT has
partially been supported by the CSIR, New Delhi, India and Indian
Institute of Technology Kanpur, Kanpur-208016, Uttar Pradesh,
India.
\section*{Appendix A}
In this appendix, we provide precise expression for the case of two
parameter thermodynamic configuration. This review set-up is offered from
the viewpoint of two parameter family giant and superstar solutions.
In order to do so, we recall important recent studies of the thermodynamic
properties of diverse (rotating) black holes have elucidated interesting
aspects of phase transitions, if any, in the state-space geometric framework
and their associated relations with the extremal black hole solutions in the
context of $\mathcal N \geq 2$ compactifications \cite{sfm1, sfm2}. It may
be argued however that the connection of such a geometric formulation to
the thermodynamic fluctuation theory of black holes requires several
modifications \cite{rup3}. The geometric formulation thus involved has
first been applied to $\mathcal  N\geq 2$ supergravity extremal black
holes in $D=4$, which arise as low energy effective field theories from
the compactifications of Type II string theories on Calabi-Yau manifolds
\cite{fgk}. Since then, several authors have attempted to understand
this connection \cite{cai1,gr-qc/0304015v1,0510139v3,Arcioni}, for both
the supersymmetric as well as non-supersymmetric four dimensional black
holes and five dimensional rotating black string and ring solutions.
Interesting discussions on the vacuum phase transitions, if any,
exist in the literature, which involves some change of the black hole
horizon topology, \cite{cai1,gr-qc/0304015v1,0510139v3,Arcioni}.

Ruppenier has conjointly advocated the assumption ``that
all the statistical degrees of freedom of black hole live on the
black hole event horizon'', and thus the scalar curvature signifies
the average number of correlated Planck areas on the event horizon
of the black hole \cite{RuppeinerPRD78}. Specifically, the zero
scalar curvature indicates certain bits of information on the
event horizon fluctuating independently of each other, while the
diverging scalar curvature signals a phase transition indicating
highly correlated pixels of the informations. Moreover, Bekenstein
has introduced an elegant picture for the quantization of the area of
the event horizon, being defined in terms of Planck areas
\cite{Bekenstein}. Recently, the state-space geometry of the
equilibrium configurations thus described has extensively been
applied to study the thermodynamics of a class of rotating black
hole configurations \cite{bnt,BNTBull,bntSb, BSBR}.

From the viewpoint of the present consideration,
the underlying moduli configuration appears to be horizonless and
smooth. However, in the classical limit in which the Planck length and
the AdS throat scale as, respectively $l_P\rightarrow 0$ and $L\rightarrow 0$,
in such a way that their ratio diverges $l_P/L \rightarrow \infty$,
the underlying moduli configuration acquires an entropy which may
be assumed to be associated with the average horizon area of the 
black hole. In general, one wishes to compare the quantization of a
classical moduli space from the known perspective of the AdS/CFT
correspondence \cite{LLM, SB}. In this concern, it is worth to
mention that we have 
\begin{scriptsize}
\begin{eqnarray}
AdS/CFT: \ \{\mathcal{N}=4 \ SYM \}\ \leftrightarrow \ \{Type \
IIB \ String \ Theory \ on \ AdS_5 \times S^5 \}
\end{eqnarray}
\end{scriptsize}
Following \cite{hep-th/0508023,hep-th/0107119}, it is important to mention
that the supergravity description emerges in the strong coupling limit with
\begin{scriptsize} $g^2_{YM}N>>$ \end{scriptsize}1%. On the other hand
while, the dual CFT emerges in the weak coupling limit \begin{scriptsize}
$g^2_{YM}N<<1$\end{scriptsize}. Thus, one finds the matching of entropies
in the $\frac{1}{2}$ BPS sector of the configuration \cite{hep-th/0508023,
hep-th/0107119}. In this sector, the authors of \cite{hep-th/0508023,hep-th/0107119}
have shown  that nearly all states look alike and they belong to the same chiral
ring quantization of the classical moduli space. It is however intriguing to note 
that the dual CFT formalism faces problems with the wave functional renormalization,
mixing of states, and in general it may feature certain other technical difficulties
as well \cite{hep-th/0508023, hep-th/0107119}. Nevertheless, none of these concerns
the present analysis, and thus we may safely divulge the probable fluctuations
over chosen equilibrium giants configurations. 

As mentioned in the introduction, the $1/2$-BPS (dual) giant configurations
are parameterized by two chemical potentials, $\lambda_1, \lambda_2$. Thus,
the fluctuations around the minima energy configuration are describe an
intrinsic Wienhold surface. 
Explicitly, let us consider the case of the two dimensional intrinsic chemical geometry,
such that the components of the metric tensor are given by \\
\begin{scriptsize}
\begin{eqnarray}
g_{ \lambda_1  \lambda_1}&=& \frac{\partial^2 E}{\partial  \lambda_1^2} \nonumber  \\
g_{ \lambda_1  \lambda_2}&=& \frac{\partial^2 E}{{\partial  \lambda_1}{\partial  \lambda_2}}\nonumber  \\
g_{ \lambda_2  \lambda_2}&=& \frac{\partial^2 E}{\partial
\lambda_2^2}
\end{eqnarray}
\end{scriptsize}
In this case, it follows that the determinant of the metric tensor is 
\begin{scriptsize}
\begin{eqnarray}
\Vert g(\lambda_1,\lambda_2) \Vert &= &S_{\lambda_1 \lambda_1}S_{\lambda_2 \lambda_2}- S_{\lambda_1 \lambda_2}^2
\end{eqnarray}
\end{scriptsize}
Now, we can calculate the $\Gamma_{ijk}$, $R_{ijkl}$, $R_{ij}$ and
$ R $ for the above two dimensional thermodynamic geometry $(M_2,g)$.
One may easily inspect that the scalar curvature is given by
\begin{scriptsize}
\begin{eqnarray}
R(\lambda_1, \lambda_2)&=& -\frac{1}{2} (E_{ \lambda_1 \lambda_1}E_{ \lambda_2
\lambda_2}- E_{ \lambda_1  \lambda_2}^2)^{-2} (E_{ \lambda_2
 \lambda_2}E_{ \lambda_1 \lambda_1 \lambda_1}E_{ \lambda_1  \lambda_2
 \lambda_2}\nonumber
\\ &&+ E_{ \lambda_1  \lambda_2}E_{ \lambda_1 \lambda_1 \lambda_2}E_{
\lambda_1 \lambda_2 \lambda_2}+
E_{ \lambda_1 \lambda_1}E_{ \lambda_1 \lambda_1 \lambda_2}E_{ \lambda_2 \lambda_2 \lambda_2}\nonumber  \\
&&- E_{ \lambda_1 \lambda_2}E_{ \lambda_1 \lambda_1 \lambda_1}E_{
\lambda_2 \lambda_2 \lambda_2}- E_{ \lambda_1 \lambda_1}E_{
\lambda_1 \lambda_2 \lambda_2}^2 - E_{ \lambda_2 \lambda_2}E_{
\lambda_1 \lambda_1 \lambda_2}^2 )
\end{eqnarray}
\end{scriptsize}
Furthermore, the relation between the chemical scalar curvature
and the Riemann covariant curvature tensor for any two dimensional
intrinsic geometry is given (see for details \cite{bnt}) by
\begin{scriptsize}
\begin{eqnarray} \label{Wiencur}
R=\frac{2}{\Vert g \Vert}R_{ \lambda_1 \lambda_2 \lambda_1 \lambda_2}
\end{eqnarray}
\end{scriptsize}
The relation Eq. \ref{Wiencur} is quite usual for an arbitrary
intrinsic Riemannian surface $(M_2(R),g)$. 
Correspondingly, the Legendre transformed version of the Wienhold manifold
is known as state-space manifold. In fact, the associated configuration
is parameterized by the number of excited and total boxes, i.e. $ \{n,M\}$. 
Thus, the Gaussian fluctuation of the entropy of $1/2$-BPS configurations
form a two dimensional state-space surface $M_2$.
Explicitly, the components of covariant state-space metric tensor may be given as
\begin{scriptsize}
\begin{eqnarray}
g_{nn}&=&- \frac{\partial^2 S(n,M)}{\partial n^2} \nonumber \\
g_{nM}&=&- \frac{\partial^2 S(n,M)}{{\partial n}{\partial M}}  \nonumber  \\
g_{MM}&=&- \frac{\partial^2 S(n,M)}{\partial M^2}
\end{eqnarray}
\end{scriptsize}
It is easy to express, in this simplest case, that the
determinant of the metric tensor turns out to be
\begin{scriptsize}
\begin{eqnarray}
\Vert g(n,M) \Vert &= &S_{nn}S_{MM}- S_{nM}^2
\end{eqnarray}
\end{scriptsize}
As in the case of Wienhold geometry, we find that the state-space scalar curvature is given by
\begin{scriptsize}
\begin{eqnarray}
R(n,M)&=& \frac{1}{2} (S_{nn}S_{MM}- S_{nM}^2)^{-2}
(S_{MM}S_{nnn}S_{nMM} + S_{nM}S_{nnM}S_{nMM} \nonumber\\ &+&
S_{nn}S_{nnM}S_{MMM} -S_{nM}S_{nnn}S_{MMM}- S_{nn}S_{nMM}^2-
S_{MM}S_{nnM}^2 )
\end{eqnarray}
\end{scriptsize}
Following the observations of \cite{BNTBull}, it is essentially evident
that the scalar curvature and the corresponding Riemann curvature tensor
of an arbitrary two dimensional intrinsic state-space manifold
$(M_2(R),g)$ may be given by
\begin{scriptsize}
\begin{eqnarray}
R(n,M)=\frac{2}{\Vert g \Vert}R_{nMnM}(n,M)
\end{eqnarray}
\end{scriptsize}
\section*{Appendix B}
In this appendix, we provide the explicit form of the most general
thermodynamic scalar curvatures describing the family of two
charged giant and superstars. Our analysis illustrates that the
physical properties of the specific scalar curvatures may exactly
be exploited, without any approximation. The definite behavior of
curvatures, as accounted in section three, suggests that the
various intriguing chemical and state-space examples of $1/2$ BPS
solutions include the nice property that they do not diverge,
except for the determinant singularity. As mentioned in the main
sections, these configurations are an interacting statistical
system. We discover that their thermodynamic geometries indicate
the possible nature of general two parameter equilibrium
configurations. Significantly, one may notice, from the very
definition of intrinsic metric tensors, that the relevant Riemann
covariant curvature tensors and scalar curvatures may be thus
presented as follows.
\subsection*{(i) Canonical Energy Fluctuations:}
Here, we shall explicitly supply the exact Riemann covariant
tensor of the Weinhold geometry for the 2- parameter giants and
superstars. It turns out that the functional nature of a large
number of gravitons, within the neighborhood of small chemical
fluctuations introduced in the canonical ensemble of $1/2$ BPS
configurations, may precisely be divulged. Surprisingly, we expose
in this framework, that the intrinsic covariant curvature tensor
takes the exact and simple expression
\begin{scriptsize}
\begin{eqnarray} \label{canenfluc}
R_{T \lambda T \lambda}(T, \lambda)&=& -\frac{1}{4}
\{(\sum_{j=0}^{\infty}(-j^2 \frac{b_j}{T^4}\frac{(-2 T+2 T b_j+j+j
b_j)} {(b_j-1)^3}))\times
(\sum_{j=0}^{\infty}(-j b_j\frac{(1+b_j)}{(b_j-1)^3}) \\
\nonumber && -(\sum_{j=0}^{\infty}(j^2 \frac{b_j}{T^2}
\frac{(1+b_j)}{(b_j-1)^3}))^2\}^{-1} \{-(\sum_{j=0}^{\infty}(-j
b_j\frac{(1+b_j)}{(b_j-1)^3})) \times \\ \nonumber &&
(\sum_{j=0}^{\infty}(-j^2 \frac{b_j}{T^4}\frac{(-2 T+2 T b_j^2+j+4
j b_j+j b_j^2)}{(b_j-1)^4}))^2 +(\sum_{j=0}^{\infty}(-j
b_j\frac{(1+b_j)}{(b_j-1)^3})) \times
\\ \nonumber && (\sum_{j=0}^{\infty}(j^2 \frac{b_j}{T^6}
\frac{(j^2+4 j^2 b_j+j^2 b_j^2-6 j T+6 T^2+6 j T b_j^2-12 T^2
b_j+6 T^2 b_j^2)}{(b_j-1)^4})) \\ \nonumber && \times
(\sum_{j=0}^{\infty}(j^2 \frac{b_j}{T^2} \frac{(1+4
b_j+b_j^2)}{(b_j-1)^4})) +(\sum_{j=0}^{\infty}(j^2
\frac{b_j}{T^2} \frac{(1+b_j)} {(b_j-1)^3})) \times \\
\nonumber && (\sum_{j=0}^{\infty}(-j^2 \frac{b_j}{T^4}\frac{(-2
T+2 T b_j^2+j+4 j b_j+j b_j^2)}{(b_j-1)^4})) \times \\
\nonumber && (\sum_{j=0}^{\infty}(j^2 \frac{b_j}{T^2}\frac{(1+4
b_j+b_j^2)}{(b_j-1)^4})) -(\sum_{j=0}^{\infty}(j^2
\frac{b_j}{T^2} \frac{(1+b_j)}{(b_j-1)^3})) \times \\
\nonumber && (\sum_{j=0}^{\infty}(j^2 \frac{b_j}{T^6} \frac{(j^2+4
j^2 b_j+j^2 b_j^2-6 j T+6 T^2+6 j T b_j^2-12 T^2 b_j+6 T^2
b_j^2)}{(b_j-1)^4})) \\ \nonumber && \times
(\sum_{j=0}^{\infty}(-j b_j \frac{(1+4 b_j+b_j^2)} {(b_j-1)^4}))
-(\sum_{j=0}^{\infty}(-j^2 \frac{b_j}{T^4}\frac{(-2 T+2 T b_j+j+j
b_j)}{(b_j-1)^3})) \times \\ \nonumber && (\sum_{j=0}^{\infty}(j^2
\frac{b_j}{T^2} \frac{(1+4 b_j+b_j^2)}{(b_j-1)^4}))^2
+(\sum_{j=0}^{\infty}(-j^2 \frac{b_j}{T^4} \frac{(-2 T+2 T b_j+j+j
b_j)} {(b_j-1)^3})) \times \\ \nonumber &&
(\sum_{j=0}^{\infty}(-j^2 \frac{b_j}{T^4} \frac{(-2 T+2 T
b_j^2+j+4 j b_j+j b_j^2)}{(b_j-1)^4})) \times (\sum_{j=0}^{\infty}
(-j b_j\frac{(1+4 b_j+b_j^2)} {(b_j-1)^4})) \}
\end{eqnarray}
\end{scriptsize}
\subsection*{(ii) State-space Fluctuations:}
As stated earlier, the state-space metric in the excited and
unexcited droplets is given by the negative Hessian matrix of the
box counting entropy. Here, the numbers of excited and empty boxes
in a given Young diagram are respected to be extensive variables.
In this case,  the two distinct large integers characterize the
intrinsic state-space correlation length, carried by the giants
and superstars. Our computation shows that the exact Riemann
covariant curvature tensor is given by
\begin{scriptsize}
\begin{eqnarray} \label{countentcovcur}
R_{nMnM}(n,M) &=&
1/2\{(-2M^2\Psi(1,n+1)\Psi(1,M^2+1)-\Psi(1,n+1)\Psi(M^2+1)
\\ \nonumber && +2M^2\Psi(1,n+1)\Psi(1,M^2-n+1) +\Psi(1,n+1)
\Psi(M^2-n+1) \\ \nonumber && -2M^2\Psi(1,M^2-n+1)
\Psi(1,M^2+1)-\Psi(1,M^2-n+1)\Psi(M^2+1) \\ \nonumber &&
+\Psi(1,M^2-n+1)\Psi(M^2-n+1))\}^{-1} [
\Psi(1,n+1)\Psi(1,M^2-n+1)^2 \\ \nonumber && -\Psi(1,M^2-n+1)
\Psi(2,n+1)\Psi(M^2+1) \\ \nonumber && +\Psi(1,M^2-n+1)\Psi(2,n+1)
\Psi(M^2-n+1) \\ \nonumber && +\Psi(2,M^2-n+1)\Psi(1,M^2-n+1)
\Psi(M^2+1)\\ \nonumber &&-\Psi(2,M^2-n+1)\Psi(1,M^2-n+1)
\Psi(M^2-n+1) \\ \nonumber && +2 M^2 \{ 3\Psi(2,M^2-n+1)
\Psi(1,n+1) \Psi(1,M^2+1) \\ \nonumber && +2\Psi(1,M^2-n+1)
\Psi(2,n+1) \Psi(1,M^2+1) \\ \nonumber &&
-2\Psi(1,M^2-n+1)^2\Psi(2,n+1) +\Psi(1,M^2-n+1)^3\\ \nonumber &&
-\Psi(2,M^2-n+1) \Psi(2,n+1)\Psi(M^2+1)\\ \nonumber &&
+\Psi(2,M^2-n+1)\Psi(2,n+1) \Psi(M^2-n+1)\\ \nonumber &&
-\Psi(2,M^2-n+1) \Psi(1,n+1)\Psi(1,M^2-n+1) \\ \nonumber &&
+\Psi(2,M^2-n+1)\Psi(1,M^2-n+1)\Psi(1,M^2+1)\} \\ \nonumber &&
+4M^4 \{ \Psi(2,M^2-n+1)\Psi(1,n+1)\Psi(2,M^2+1)\\ \nonumber &&
-\Psi(2,M^2-n+1)\Psi(2,n+1) \Psi(1,M^2+1) \\ \nonumber &&
+\Psi(1,M^2-n+1)\Psi(2,n+1) \Psi(2,M^2+1)\}]
\end{eqnarray}
\end{scriptsize}
Finally, let us turn our attention to the state-space scalar
curvature for the two parameter droplet configurations. A
systematic examination demonstrates that the scalar curvature is
\begin{scriptsize}
\begin{eqnarray} \label{countentscacur}
R(n,M) &=& \frac{1}{2} \{-2 \Psi(1,n+1) \Psi(1,M^2+1)
M^2-\Psi(1,n+1) \Psi(M^2+1) \\ \nonumber && +2 M^2 \Psi(1,n+1)
\Psi(1,M^2-n+1) +\Psi(1,n+1) \Psi(M^2-n+1) \\ \nonumber && -2 M^2
\Psi(1,M^2-n+1) \Psi(1,M^2+1)- \Psi(1,M^2-n+1) \Psi(M^2+1)\\
\nonumber && +\Psi(1,M^2-n+1) \Psi(M^2-n+1)\}^{-2} \\ \nonumber
&&[ \Psi(1,M^2-n+1)^3 + \Psi(1,n+1) \Psi(1,M^2-n+1)^2 \\ \nonumber
&& -\Psi(1,M^2-n+1) \Psi(2,n+1) \Psi(M^2+1) \\
\nonumber && +\Psi(1,M^2-n+1) \Psi(2,n+1) \Psi(M^2-n+1) \\
\nonumber && +\Psi(2,M^2-n+1) \Psi(1,M^2-n+1) \Psi(M^2+1) \\
\nonumber && -\Psi(2,M^2-n+1) \Psi(1,M^2-n+1) \Psi(M^2-n+1) \\
\nonumber && +2 M^2 \{ \Psi(2,M^2-n+1) \Psi(1,M^2-n+1)
\Psi(1,M^2+1) \\ \nonumber && - \Psi(2,M^2-n+1) \Psi(1,n+1)
\Psi(1,M^2-n+1) \\ \nonumber && - \Psi(2,M^2-n+1) \Psi(2,n+1)
\Psi(M^2+1) \\ \nonumber && + \Psi(2,M^2-n+1)
\Psi(2,n+1)\Psi(M^2-n+1)\\ \nonumber && - 2 \Psi(1,M^2-n+1)^2
\Psi(2,n+1) \\ \nonumber && + 2 \Psi(1,M^2-n+1) \Psi(2,n+1)
\Psi(1,M^2+1) \\ \nonumber && +3 \Psi(2,M^2-n+1) \Psi(1,n+1)
\Psi(1,M^2+1)\} \\ \nonumber && +4 M^4 \{ \Psi(2,M^2-n+1)
\Psi(1,n+1) \Psi(2,M^2+1)\\ \nonumber && + \Psi(1,M^2-n+1)
\Psi(2,n+1) \Psi(2,M^2+1) \\ \nonumber && - \Psi(2,M^2-n+1)
\Psi(2,n+1) \Psi(1,M^2+1) \}]
\end{eqnarray}
\end{scriptsize}

\end{document}